\documentclass[11pt,a4paper]{article}
\usepackage{jheppub}
\usepackage{bm}
\usepackage{amsmath}
\usepackage{amssymb}
\usepackage{graphicx}
\usepackage[caption=false]{subfig}
\usepackage{mathtools}
\usepackage{placeins}
\usepackage{booktabs}
\usepackage{afterpage}
\usepackage{tikz-cd}
\usepackage{tikz}
\usetikzlibrary{fit,arrows.meta,matrix,positioning,backgrounds}
\pgfdeclarelayer{background}
\pgfsetlayers{background,main}

\hypersetup{urlcolor=blue,citecolor=blue,linkcolor=blue}

\newcommand{\be}{\begin{equation}}
\newcommand{\ee}{\end{equation}}
\newcommand{\half}{\textstyle{1\over 2}}
\newcommand{\quarter}{\textstyle{1\over 4}}
\newcommand{\sq}{\textstyle{1\over \sqrt{2}}}

\title{Hopf Fibration and Monopoles in the Standard Model}

\author[a]{Richard A. Battye}

\affiliation[a]{Jodrell Bank Centre for Astrophysics, Department of Physics and Astronomy, University of Manchester, Oxford Road, Manchester M13 9PL, United Kingdom}

\emailAdd{Richard.Battye@manchester.ac.uk}

\abstract{By writing the angular part of the complex doublet in terms of a ${\mathbb C}P^1$-spinor and a phase, we carefully demonstrate the Hopf fibration of the bosonic sector of the electroweak theory. This allows us to write it in terms of the modulus of the Higgs field, a unit vector field ${\bf{\hat n}}\in S^2$ and a fibre phase $\zeta\in S^1$. We define the fibre connection and curvature, the covariant derivatives of the new fields and demonstrate their gauge invariance. The first Chern number of the geometry is equivalent to the monopole number and there is a linking number associated with the Hopf invariant. We decompose the gauge fields using Cho's restricted ansatz and ultimately write the EM, Z and W field tensors in terms of the fibre curvature. We show how the standard perturbative SM can be accommodated in this framework, but also how EFT might be formulated in a different way, and that the fibration can also be applied to the fermionic sector. We show that there is no regular finite-energy isolated monopole carrying non-zero asymptotic electromagnetic magnetic charge in the unmodified SM, and identify the underlying obstruction with the non-extension of the electromagnetic transition function through the full electroweak gauge bundle. We also construct a configuration with Hopf invariant one which corresponds to a well-separated one-parameter family of monopole/antimonopole loops in $S^3$. We argue that all constructions are particularly clear within the Hopf reformulation.}

\begin{document}
\maketitle
\flushbottom

\section{Introduction}

The Standard Model (SM) of particle physics has endured significant experimental scrutiny, but most of this has been in the perturbative regime where it can accurately predict the nature of particle interactions in colliders. The global structure of the model is relevant to the existence of topological defects~\cite{VilenkinShellard2000} and solitons~\cite{MantonSutcliffe2004}, but there is very little experimental evidence to be sure that this is what is usually assumed. Conventionally, the vacuum manifold is ${\cal M}=S^3$ which supports the existence of unstable topological defects known as Sphalerons~\cite{Manton1983,KlinkhamerManton1984} due to homotopy group $\pi_3(S^3)=\mathbb{Z}$. This also predicts that there are no stable topological defects in the SM. The aim of this paper is to discuss some aspects of the global structure of the SM and clarify the possibility of 3D topological defects.

The most obvious of these are monopoles (see \cite{Rajantie:2016paj} for a high level review and also \cite{GoddardOlive1978} and \cite{Preskill1984} for historical perspectives) which are being searched for by the MoEDAL experiment~\cite{MoEDAL:2019ort}. They also could play a role in the generation of primordial magnetic fields~\cite{Vachaspati:1991nm,Vachaspati:2020blt} if they are formed in the Early Universe. However, there are no known solutions in the SM. 

The electroweak theory (EWT), part of the SM, is usually formulated in terms of a complex doublet $\Phi\in\mathbb{C}^2$ with bosonic Lagrangian density  
\be 
{\cal L}_B=|D_\mu\Phi|^2-V(\Phi)-\quarter Y_{\mu\nu}Y^{\mu\nu}-\quarter{\bf W}_{\mu\nu}\cdot{\bf W}^{\mu\nu}\,, 
\label{smlag}
\ee
where $D_\mu\Phi=(\partial_\mu+\half ig{\bf W}_\mu\cdot\boldsymbol{\sigma}+\half ig^{\prime}Y_\mu)\Phi$. The field strength tensors for the $U(1)_Y$ and $SU(2)_L$ fields are given by $Y_{\mu\nu}=\partial_\mu Y_\nu-\partial_\nu Y_\mu$ and ${\bf W}_{\mu\nu}=\partial_\mu{\bf W}_\nu-\partial_\nu{\bf W}_\mu-g{\bf W}_\mu\times{\bf W}_\nu$ and the potential is $V(\Phi)=\lambda(|\Phi|^2-\half v_{\rm SM}^2)^2$. We will be concerned mainly with the EW part of the SM. The fermionic part of the Lagrangian, ${\cal L}_F$, and the Yukawa coupling Lagrangian, ${\cal L}_Y$ are  discussed in section~\ref{sec:fermion} and we will ignore the $SU(3)$ part of the SM. 

The objective of this paper is to derive an exactly equivalent formulation of the EWT in terms of the Hopf fibration~\cite{Hopf1931}\footnote{A fibre bundle is a map $\pi:E\rightarrow B$ for which $\pi^{-1}(b)=F$ for all $b\in B$. $B$ is known as the base space and $F$ is the fibre. The Hopf fibration has $(E,B,F)=(S^3,S^2,S^1)$.}. We will see that the Lagrangian can be written, without any restrictions except that $|\Phi|\ne 0$, in terms of a unit vector
\be 
{\hat n}^a=-{\Phi^\dag\sigma^a\Phi\over\Phi^\dag\Phi}\,,
\ee
and a fibre phase, $\zeta$. We will demonstrate how the modified scalar kinetic term respects the same  gauge invariance as (\ref{smlag}). In particular, if we write $\Phi=\sq v U{\bf e}$ where $U\in SU(2)_L$ and ${\bf e}\in {\mathbb C}^2$ is constant, then ${\hat n}^a={\cal R}^{ab}{\bar n}^b$ where ${\bar n}^a=-{\bf e}^\dag\sigma^a{\bf e}$ is ${\hat z}^a$ for the standard choice of ${\bf e}^T=(0\quad 1)$. Using a particular form for $U=\cos\half\gamma\,\sigma^0+i\sin\half\gamma\,{\hat{\bf G}}\cdot{\boldsymbol{\sigma}}$, where $0\le\gamma\le 2\pi$ and $|{\hat{\bf G}}|^2=1$, then 
\be 
{\cal R}^{ab}(\gamma,{\hat{\bf G}})={\hat G}^a{\hat G}^b+\cos\gamma (\delta^{ab}-{\hat G}^a{\hat G}^b)+\sin\gamma\varepsilon^{abc}{\hat G}^c\,,
\ee
which is the $SO(3)$ matrix for a rotation of $\gamma$ around the axis ${\hat{\bf G}}$. It will also be helpful to define $U^\dag\partial_\mu U=\half igL_\mu^a\sigma^a$ where
\be 
L_\mu^a(\gamma,{\hat{\bf G}})={1\over g}\left[\partial_\mu\gamma\,{\hat G}^a+\sin\gamma\,\partial_\mu{\hat G}^a+(1-\cos\gamma)\varepsilon^{abc}\,{\hat G}^b\partial_\mu{\hat G}^c\right]\,,
\label{lia}
\ee
is sometimes known as the Maurer-Cartan vector. Setting $\gamma=0$ is the case of unitary gauge.

The unit vector field ${\bf{\hat n}}({\bf x})$ has  clear physical meaning: it is the longitudinal direction in field space that defines the photon and Z boson field. In particular, we have that  
\be
A_\mu=\sin\theta_W\,{\bf W}_\mu\cdot{\hat{\bf n}}+\cos\theta_W\,Y_\mu\,,\quad Z_\mu=\cos\theta_W\,{\bf W}_\mu\cdot{\hat{\bf n}}-\sin\theta_W\,Y_\mu\,,
\label{azdefn}
\ee
where $\tan\theta_W=g^{\prime}/g$. We will see that, within the vacuum manifold $S^3$, the fibre phase, $\zeta({\bf x})$, corresponds to rotation about the ${\bf{\hat n}}$ axis and that it is connected with the hypercharge symmetry. 

Expressing the EWT in terms of ${\bf{\hat n}}$ is not a new suggestion and there are a number of works which have attempted to reformulate the EWT and other gauge theories in this way. The first was the restricted gauge theory of Cho~\cite{Cho1980} which we will discuss in section~\ref{sec:cho}. Faddeev and Niemi (FN) suggested an extension of this idea in the context of $SU(2)$ Yang-Mills theory~\cite{FaddeevNiemi1999}.  A more directly related reformulation of the bosonic electroweak theory was subsequently given by Chernodub, Faddeev and Niemi~\cite{Chernodub:2008bd}, who introduced gauge-invariant supercurrent variables together with a local moving frame defined by the normalized Higgs doublet. Their construction contains an internal $U(1)$ redundancy which retains the full angular information of the Higgs field, and Faddeev~\cite{Faddeev:2008} later gave an equivalent formulation emphasizing the $SU(2)\cong S^3$ structure of the normalized doublet. These constructions contain several of the local ingredients used below. Our emphasis is that the fibre phase is necessary to maintain the correct number of degrees of freedom and gauge invariance. In addition it also plays a significant role in the understanding of the geometrical nature of the reformulation - this will be explained in section~\ref{sec:geom}.

Having written the EWT in terms of ${\bf{\hat n}}$, a natural question is to ask is whether it can support a monopole solution and if so can these lead to a modification to Maxwell's laws. The study of monopole-like configurations\footnote{Here, and in the rest of the paper we will use the word {\em solution} to refer to a regular, finite-energy solution with no divergences. By that definition all the previously suggested monopoles are {\em configurations} as opposed to solutions.} in the EWT has a long history~\cite{Dirac1931,WuYang1975,WuYang1976,Nambu1977,ChoMaison1997}, but no finite-energy solution has been found so far. In section~\ref{sec:mono} we first show that there are no regular finite energy spherically symmetric monopole solutions and then show that a more general gauge-bundle obstruction excludes a regular isolated monopole carrying non-zero asymptotic EM magnetic charge.

\section{Hopf fibration of the scalar kinetic term}
\label{sec:hopf}

In this section we take the scalar kinetic term $|D_\mu\Phi|^2$ and rewrite it in terms of field bilinears. We will see that when it is done correctly, the number of degrees of freedom and gauge invariance are manifest. This is at the expense of the introduction of the fibre phase, sometimes known as the Berry phase~\cite{Berry1984,Simon1983}.

\subsection{Extraction of a phase : $\mathbb{C}P^1$ fibration}
\label{sec:phase}

Let us start with a ${\mathbb{C}}^2$ field and write it 
\be
\Phi=\begin{pmatrix}A_1e^{i\alpha_1}\cr A_2e^{i\alpha_2}\end{pmatrix}=\sqrt{A_1^2+A_2^2}e^{i\alpha_1}\begin{pmatrix}{A_1\over\sqrt{A_1^2+A_2^2}}\cr{{A_2\over\sqrt{A_1^2+A_2^2}}e^{i(\alpha_2-\alpha_1)}}\end{pmatrix}\,,
\ee
where $A_1$, $A_2\in\mathbb{R}$, and $\alpha_1$, $\alpha_2$ are phases $\in (0,2\pi]$. In the case where both of $A_1$ and $A_2$ are not zero, we can write this as 
\be 
\Phi={1\over\sqrt{2}}ve^{\half i\zeta}\xi\,,
\ee
where $\xi$ is a representative of  $\mathbb{C}P^1$ with the phase redundancy discussed below, that is, $\xi^\dag\xi=1$ which also implies that $\xi^\dag\partial_\mu\xi+(\partial_\mu\xi)^\dag\xi=0$. In this parameterization there is  1 degrees-of-freedom (dof) in $v>0$, which is the modulus of the Higgs that becomes the vacuum expectation value (VEV) in the vacuum, 1 in $\zeta$ which is the fibre phase and 2 in $\xi\in{\mathbb{C}}P^1\cong S^2$, and we have made a transformation from a field with a target space ${\mathbb{C}}^2$ to ${\mathbb{R}}_+\times S^1\times S^2$. Note that stable topological solutions in 3D  are defined by mappings from $S^2\rightarrow{\cal M}$ and solutions with one unstable mode by mappings from $S^3\rightarrow{\cal M}$. Using this change of variables, the covariant derivative of $\Phi$ is 
\begin{eqnarray} 
D_\mu\Phi
&=&{1\over\sqrt{2}}e^{\half i\zeta}\left[v\partial_\mu\xi+\left(\partial_\mu v+{1\over 2} iv\left(\partial_\mu\zeta+gW_\mu^a\sigma^a+g^{\prime}Y_\mu\right)\right)\xi\right]\,,
\end{eqnarray} 
and one can define the unit vector ${\hat n}^a=-\xi^\dag\sigma^a\xi$.

There are two important relations between the components of Pauli matrices 
\be
\sigma_{ij}^a\sigma_{kl}^a=2\delta_{il}\delta_{jk}-\delta_{ij}\delta_{kl}\,,\quad\epsilon^{abc}\sigma_{ij}^b\sigma_{kl}^c=i\left[\sigma_{il}^a\delta_{jk}-\sigma_{kj}^a\delta_{il}\right]\,, 
\label{complete}
\ee
and these allow us to deduce that    
\begin{eqnarray}
(\partial_\mu{\bf{\hat n}})^2&=&4\left[\partial_\mu\xi^\dag\partial^\mu\xi+(\xi^\dag\partial_\mu\xi)^2 \right]\,,\cr
\varepsilon^{abc}{\hat n}^b\partial_\mu{\hat n}^c&=&i\left[\xi^\dag\sigma^a\partial_\mu\xi-(\partial_\mu\xi)^\dag\sigma^a\xi+2(\xi^\dag\partial_\mu\xi){\hat n}^a\right]\,.
\end{eqnarray}
Now define $\xi_\perp$ such that $\xi^\dag\partial_\mu\xi_\perp=0$ allowing us to write 
$\partial_\mu\xi=(\xi^\dag\partial_\mu\xi)\xi+\partial_\mu\xi_\perp$, in which case one finds that 
\begin{eqnarray}
(\partial_\mu{\bf{\hat n}})^2=4 \partial_\mu\xi^\dag_\perp\partial^\mu\xi_\perp\,,\quad 
\varepsilon^{abc}{\hat n}^b\partial_\mu{\hat n}^c=i\left[\xi^\dag\sigma^a\partial_\mu\xi_\perp-(\partial_\mu\xi_\perp)^\dag\sigma^a\xi\right]\,.
\end{eqnarray}

The covariant derivative can be written as 
\be 
D_\mu\Phi={1\over\sqrt{2}}e^{\half i\zeta}\left[v\partial_\mu\xi_\perp+\left(\partial_\mu v +{1\over 2}ivQ_\mu\right)\xi\right]\,,
\label{oldcov}
\ee
where $Q_\mu=(\partial_\mu\zeta+a_\mu+ g^\prime Y_\mu)\sigma^0+gW_\mu^a\sigma^a$ and $a_\mu=-2i\xi^\dag\partial_\mu\xi$ is the U(1) (or fibre) connection.
Hence, we deduce that  
\be
|D_\mu\Phi|^2=\half\left(\partial_\mu v)^2+\textstyle{1\over 8}v^2[(D_\mu{\hat{\bf n}})^2+(D_\mu\zeta)^2\right]\,,
\label{final}
\ee
and we have defined the covariant derivatives of the bilinear and fibre phase as 
\begin{eqnarray}
D_\mu{\hat n}^a&=&\partial_\mu{\hat n}^a-g\varepsilon^{abc}W_\mu^b{\hat n}^c\,,\cr
D_\mu\zeta&=&\partial_\mu\zeta+a_\mu+g^{\prime}Y_\mu-gW_\mu^a{\hat n}^a=c_\mu+a_\mu-gW_\mu^a{\hat n}^a=\partial_\mu\zeta+a_\mu-g_ZZ_\mu\,,
\end{eqnarray}
where $g_Z=\sqrt{g^2+(g^{\prime})^2}$ and $c_\mu=\partial_\mu\zeta+g^{\prime}Y_\mu$ which will be used in section~\ref{sec:fermion}. This rewriting of the $|D_\mu\Phi|^2$ is one of the main results of this paper.

We have replaced the Higgs field with $v$, ${\hat{\bf n}}$ and $\zeta$ plus the fibre phase/connection. There are the same number of degrees of freedom. The perturbations of $v$ around the VEV $v=v_{\rm SM}$ represent the Higgs particle, with ${\bf{\hat n}}$ and $\zeta$ being the gauge degrees of freedom which can be fixed in a perturbative analysis see section~\ref{sec:pert}. The original target space was ${\mathbb{C}}^2$ and we have replaced this locally with ${\mathbb{R}}_+\times {\mathbb{C}}P^1_{\hat{\bf n}}\times S^1_\zeta$ where, of course ${\mathbb{C}}P^1\cong S^2$. It is important that we have excluded the point $\Phi=0$ and also that this is not a full direct product. 

The bilinear field ${\hat{\bf n}}$ is a mapping from $S^3\rightarrow S^2$. The preimage of the neighbourhoods $U_i\in S^2$ are $\cong U_i\times S^1$, but the maps on the overlap $U_i\cap U_j$ for $i\ne j$ can be different. This is a special case of the fibre bundle $S^1\rightarrow S^{2N+1}\rightarrow {\mathbb{C}}P^{N}$ for $N=1$ and it is equivalent to the Hopf fibration, sometimes known as the first Hopf fibration or the complex Hopf fibration. Expressed explicitly we are using the $U(1)$ fibre action $(z_1,z_2)\rightarrow e^{i\theta}(z_1,z_2)$ which maintains $|z_1|^2+|z_2|^2=1$, rather than the actual Hopf map, but geometrically they are the same. In section~\ref{sec:fibhopf}, we will repeat the calculation in the more standard Hopf coordinates.  

\subsection{Gauge invariance}

The kinetic term of the EWT is invariant under 
\be
\Phi\rightarrow e^{\half i\alpha}\Phi\,,\quad  Y_{\mu}\rightarrow Y_{\mu}-{1\over g^{\prime}}\partial_\mu\alpha\,\hbox{ and }\,
\Phi\rightarrow U\Phi\,,\quad W_\mu\rightarrow U W_\mu U^\dag-(\partial_\mu U)U^\dag\,,
\ee
where $U\in SU(2)$ and $W_\mu=\half i g{\bf W}_\mu\cdot\boldsymbol{\sigma}$. The first of these is the hypercharge transformation and can be trivially understood as part of the fibre transformation $\zeta\rightarrow\zeta+\alpha$ and the invariance of $\partial_\mu\zeta+g^\prime Y_\mu$ and $Y_{\mu\nu}$. The bilinear field ${\bf{\hat n}}$ is neutral under the hypercharge transformation.
 
The second is a little more complicated. We have that ${\hat n}^a\rightarrow {\cal R}^{ab}{\hat n}^b$, that is, it transforms under an $SO(3)$ rotation with $U$ and $-U$ corresponding to same transformation - the well-known double cover.  The transformation of $W_\mu^a$ corresponds to 
\be 
W_\mu^a\rightarrow {\cal R}^{ab}\left(W_\mu^b-L_\mu^b\right)\,,
\ee
and the transformation $\Phi\rightarrow U\Phi$ leads to $\xi\rightarrow U\xi$. Therefore, we can show that the fibre connection transforms according to 
\be 
a_\mu\rightarrow a_\mu-gL_\mu^a{\hat n}^a\,,
\ee
so that the combination $a_\mu-gW_\mu^a{\hat n}^a$ is invariant under the gauge transformations. One can show that $D_\mu{\hat n}^a={\cal R}_U^{ab}D_\mu{\hat n}^b$ using $\varepsilon^{abc}={\cal R}_U^{ad}{\cal R}_U^{be}{\cal R}_U^{cf}\varepsilon^{def}$ and $\partial_\mu{\cal R}^{ab}_U=-g{\cal R}_U^{ac}\varepsilon^{cdb}L_\mu^d$. The second of these is derived by differentiating $U^\dag\sigma^aU={\cal R}^{ab}_U\sigma^b$ and using $\partial_\mu U=UL_\mu$, along with $\partial_\mu U^\dag=-L_\mu U^\dag$. In addition, we can show that $W_{\mu\nu}^a$ also transforms under an $SO(3)$ rotation, that is $W_{\mu\nu}^a\rightarrow {\cal R}^{ab}W_{\mu\nu}^b$. 

We can see that, if one takes into account the fibre phase, the EW Lagrangian written in terms of ${\bf{\hat n}}$ and $\zeta$ is gauge invariant. However, the fibre phase and connection are absolutely crucial to this. Without them it would be impossible for the theory to remain gauge invariant.

\subsection{Fibration using Hopf coordinates}
\label{sec:fibhopf}

In section~\ref{sec:phase} we reformulated the EWT by extracting a complex phase and using a $\mathbb{C}P^1$ spinor. This is the Hopf fibration because $\mathbb{C}P^1\cong S^2$ - fibrations are geometrical and are not wedded to a specific set of coordinates. In this section we will show that one gets exactly the same result using the more traditional Hopf coordinates. We do this by writing $\Phi={1\over\sqrt{2}}v\,U_NU_\chi{\bf e}$ from which we can deduce that ${\hat n}^a=-{\bf e}^\dag U_\chi^\dag (U_N^\dag\sigma^a U_N) U_\chi{\bf e}$. We define $\chi$ such that $U_\chi{\bf e}=A{\bf e}$ where $A\in{\mathbb{C}}$ and $|A|=1$ so that ${\hat n}^a=-{\bf e}^\dag(U_N^\dag\sigma^a U_N){\bf e}=-{\cal R}^{ab}_N{\bf e}^\dag\sigma^b{\bf e}$. If we define ${\bar n}^a=-{\bf e}^{\dag}\sigma^a{\bf e}$ then ${\hat n}^a={\cal R}^{ab}_N{\bar n}^b$ and ${\cal R}^{ab}_\chi{\bar n}^b={\bar n^a}$. Without loss of generality we can then define $U_N$ to only include rotations which are perpendicular to ${\bar n}$ and, hence, 
\be 
{\cal R}_\chi^{ab}={\bar n}^a{\bar n}^b+\cos\chi\,(\delta^{ab}-{\bar n}^a{\bar n}^b)+\sin\chi\,\varepsilon^{abc}{\bar n}^c\hbox{ 
and }U_\chi=\cos\half\chi\,\sigma^0+i\sin\half\chi\,{\bar{\bf n}}\cdot{\boldsymbol\sigma}\,.
\ee
This corresponds to the standard formulation of the Hopf map $(z_1,z_2)\rightarrow (2z_1z_2^*,|z_1|^2-|z_2|^2)$ if ${\bf e}^\dag=\begin{pmatrix}0&1\end{pmatrix}$ when the non-standard sign of ${\bf{\hat n}}$ is taken into account.

Using the definitions of section~\ref{sec:phase}, we have that  $\xi=e^{-\half i\zeta}U_NU_\chi{\bf e}$ which implies that 
\be
a_\mu=-2i\xi^\dag\partial_\mu\xi=-2i{\bf e}^\dag\left[-\half i\partial_\mu\zeta+U_N^\dag\partial_\mu U_N+U_\chi^\dag\partial_\mu U_\chi\right]{\bf e}\,.
\ee
In addition, ${\bf e}^\dag U_\chi^\dag\partial_\mu U_\chi{\bf e}=-\half i\partial_\mu\chi$ and if we write $U_N^\dag\partial_\mu U_N=\half i gL_\mu^a\sigma^a$ then $a_\mu=-\partial_\mu\left(\chi+\zeta\right)-gL^a_\mu{\bar n}^a$ and $ \partial_\mu\zeta+a_\mu=-(\partial_\mu\chi+gL_\mu^a{\bar n}^a)$ indicating that this transformation is equivalent in terms of gauge invariance.

If we now write $\Phi={1\over\sqrt{2}}v\,U_\chi\Xi$ then 
\begin{eqnarray}
D_\mu\Phi&=&{1\over\sqrt{2}}U_\chi\left[\left(\partial_\mu v+\half iv[\partial_\mu\chi{\bar n}^a\sigma^a+g{\cal R}_\chi^{ab}W_\mu^a\sigma^b+g^\prime Y_\mu]\right)\Xi+v\partial_\mu\Xi\right]\,,
\label{newcov}
\end{eqnarray}
and $\Xi=U_\chi^\dag U_N U_\chi{\bf e}=e^{\half i\zeta}U_\chi^\dag\xi$; this demonstrates that $\Xi$ and $\xi$ are equivalent $\mathbb{C}P^1$ spinors. Defining
\be
W_\mu^{\prime a}=R_\chi^{ba}W_\mu^b+{1\over g}(\partial_\mu\chi)\bar n^a\,,\qquad{\tilde n}^a=-\Xi^\dagger\sigma^a\Xi\,,\qquad{\tilde a}_\mu=-2i\Xi^\dagger\partial_\mu\Xi\,,
\ee
the covariant derivative (\ref{newcov}) has precisely the same form as (\ref{oldcov}), but with $\xi$, $a_\mu$ and $W_\mu^a$ replaced by $\Xi$, ${\tilde a}_\mu$ and $W_\mu^{\prime a}$, respectively. Hence,
\be
|D_\mu\Phi|^2={1\over2}(\partial_\mu v)^2+{1\over8}v^2\left[({\cal D}_\mu{\tilde{\bf n}})^2+\left({\tilde a}_\mu+g'Y_\mu
-gW_\mu^{\prime a}{\tilde n}^a\right)^2\right],
\label{newcovsq}
\ee
where ${\cal D}_\mu{\tilde n}^a=\partial_\mu{\tilde n}^a-g\epsilon^{abc}W_\mu^{\prime b}{\tilde n}^c$. Since ${\hat n}^a=R_\chi^{ab}{\tilde n}^b$, one finds $(D_\mu{\hat{\bf n}})^2=({\cal D}_\mu{\tilde{\bf n}})^2$ and also 
\be
{\tilde a}_\mu+g'Y_\mu-gW_\mu^{\prime a}{\tilde n}^a=\partial_\mu\zeta+a_\mu+g'Y_\mu-gW_\mu^a{\hat n}^a=D_\mu\zeta\,,
\ee
Consequently, (\ref{newcovsq}) reduces exactly to (\ref{final}).

\subsection{What happens if you ignore the fibre phase?}
\label{sec:noberry}

In section~\ref{sec:phase} and \ref{sec:fibhopf} we have demonstrated the Hopf fibration, using two different ways to extract the fibre phase from the complex doublet. In this section we will discuss what happens if one does not include this as a dynamical degree of freedom (as was the case in ref.~\cite{Cho1980}). In particular we will write $\Phi=v\xi/\sqrt{2}$. We have already pointed out that maintaining gauge invariance requires the inclusion of the fibre phase (and connection) as a dynamical field. By construction the relations that we derive in this section only apply when $\zeta=0$ and not more generally.

Using (\ref{complete}) we find that 
\be
({\hat{\bf n}}\cdot{\boldsymbol{\sigma}})\xi=-\xi\,,\quad\hbox{and}\quad [({\hat{\bf n}}\times\partial_\mu{\hat{\bf n}})\cdot{\boldsymbol{\sigma}}]\xi=2i\partial_\mu\xi+a_\mu\xi\,,
\ee
the first of which looks like an eigenvalue problem for the spinor, $\xi$. In addition, we will write the $SU(2)$ gauge field as
\be 
{\bf W}_\mu=w_\mu{\hat{\bf n}}+{1\over g}{\hat{\bf n}}\times\partial_\mu{\hat{\bf n}}+{\widetilde{\bf X}}_\mu\,,
\label{chosplit}
\ee
where ${\widetilde{\bf X}}_\mu\cdot{\hat{\bf n}}=0$. This is Cho's decomposition of the gauge fields~\cite{Cho1980}, that we will discuss in section~\ref{sec:cho}. Using these relations we can write the covariant derivative as 
\be
D_\mu\Phi={1\over\sqrt{2}}\left[\left(\partial_\mu v+\half iva_\mu- \half i gvw_\mu+\half ig^\prime vY_\mu \right)\sigma^0+\half igv{\widetilde{\bf X}}_\mu\cdot{\boldsymbol{\sigma}}\right]\xi\,,
\ee
and hence 
\be
|D_\mu\Phi|^2={1\over 2}(\partial_\mu v)^2+{1\over 8}v^2(a_\mu-g_ZZ_\mu)^2+{1\over 8}g^2v^2{\widetilde{\bf X}}_\mu\cdot{\widetilde{\bf X}}^\mu\,.
 \ee
 Setting $\zeta\equiv 0$ is a local choice of Hopf frame and does not by itself discard the fibre information, since this information remains encoded in the phase of $\xi$ and hence in $a_\mu$. A formulation purely in terms of $v$ and ${\hat{\bf n}}$, however, does not retain this information unless the fibre connection is also specified. Thus, the fibre connection is precisely the remnant of the fibre degree of freedom when the explicit coordinate $\zeta$ is fixed.

\section{Geometrical aspects of the formulation}
\label{sec:geom}

In the previous section we have reformulated the scalar kinetic term in the EWT. We noted that key aspects of this reformulation are the fibre phase and connection. Here, we discuss the important geometrical aspects of this reformulation and explain how they relate to topological invariants.

\subsection{Fibre phase, connection and curvature}
The fibre connection exists because the choice of $\zeta$ is arbitrary under the transformation $\zeta\rightarrow\zeta+\lambda$ and $\xi\rightarrow e^{-\half i\lambda}\xi$, and the quantity $\partial_\mu \zeta+a_\mu$ is invariant under this transformation. If we choose the specific form 
\be
\xi=\begin{pmatrix}-e^{-i\phi}\sin\half\theta\cr\cos\half\theta\end{pmatrix}\,, 
\label{Nambu}
\ee
then $a_\mu=-(1-\cos\theta)\partial_\mu\phi$ which is indicative of a monopole. In fact this particular, coordinate choice for the $\mathbb{C}P^1$ spinor is that of the Nambu monopole - see section~\ref{sec:nambu}. It is helpful to define the curvature of the fibre field by
\be
f_{\mu\nu}=\partial_\mu a_\nu-\partial_\nu a_\mu=-2i\left[(\partial_\mu\xi)^\dag\partial_\nu\xi-(\partial_\nu\xi)^\dag\partial_\mu\xi\right]\,,
\ee
which is given by $f_{\mu\nu}=-\sin\theta\left(\partial_\mu\theta\partial_\nu\phi-\partial_\nu\theta\partial_\mu\phi\right)$,
for the case of (\ref{Nambu}) and the corresponding ``magnetic field'' is given by $b_i=-{1\over 2}\varepsilon_{ijk}f_{jk}=\epsilon_{ijk}{\hat\theta}_j{\hat\phi}_k/(r^2)={\hat r}_i/(r^2)$ which is that of a monopole.

One can calculate the equivalent of the Aharonov-Bohm phase by integrating the fibre connection around a closed curve at fixed $\theta$. In the particular coordinate system above, this is given by the holonomy~\cite{Berry1984,Simon1983}
\be
\Delta\phi_{\rm AB}={q\over\hbar}\int a_\mu dx^\mu=-{2q\pi\over\hbar}(1-\cos\theta)\,,
\ee
which is $-2\pi q/\hbar$ for the equator $\theta=\pi/2$.

One can define the projector $P=\xi\xi^{\dag}=\half(I-{\hat{\bf n}}\cdot{\boldsymbol{\sigma}})$ and its complement as ${\tilde P}=I-P=\half(I+{\hat{\bf n}}\cdot{\boldsymbol{\sigma}})$ and, therefore,  we can write $(\partial_\mu\xi)^\dag\partial_\nu\xi=(\partial_\mu\xi)^\dag(P+{\tilde P})\partial_\nu\xi$. Using this relation we can deduce that 
\be 
f_{\mu\nu}= i{\hat n}^a[(\partial_\mu\xi)^\dag\sigma^a\partial_\nu\xi-(\partial_\nu\xi)^\dag\sigma^a\partial_\mu\xi]=-{\bf{\hat n}}\cdot\partial_\mu{\hat {\bf n}}\times\partial_\nu{\hat {\bf n}}\,,
\label{chern1}
\ee
which is the  pullback of the area two form on $S^2$.

The first Chern number, $c_1$ is a measure of the obstruction due to `twisting' of  $S^3$ being written as a direct product $S^2\times S^1$. Using (\ref{Nambu}) one can calculate
\be 
c_1=-{1\over 8\pi}\int_{S^2}d\theta d\phi\,\varepsilon^{ij}f_{ij}={1\over 4\pi}\int^\pi_0 d\theta\int_0^{2\pi}d\phi\sin\theta\,\varepsilon^{ij}\partial_i\theta\partial_j\phi=1\,,
\label{chern2}
\ee
which indicates that for the choice (\ref{Nambu}), there is no universal patch on which the mapping is one-to-one - this is the same patching structure as in the Wu and Yang monopole~\cite{WuYang1975,WuYang1976} realized here for the Nambu Monopole~\cite{Nambu1977}. For choices where $c_1$ is zero, then there would be a trivial one-to-one mapping and there is a global section.  

\subsection{Topological invariants}

If we work within the sector of the EWT where $c_1\ne 0$ then we enforce the Hopf fibration upon the Higgs field. There is a topological winding number associated with the unit vector ${\hat{\bf n}}$ which corresponds to the monopole number
\be
N_{\hat{\bf n}}={1\over 8\pi}\int dS_i\,\varepsilon^{ijk}\varepsilon^{abc}{\hat n}^{a}\partial_{j}{\hat n}^{b}\partial_{k}{\hat n}^{c}\,,
\ee 
which is an integral over a surface at infinity and we have that  $N_{\bf{\hat n}}=c_1$ using (\ref{chern1}) and (\ref{chern2}). This is the charge associated with the homotopy group $\pi_2(S^2)={\mathbb{Z}}$.

It is important to distinguish the $U(1)$ bundle appearing in the Hopf lift of ${\hat{\bf n}}$ from the physical EW gauge bundle. A non-zero $c_1$ implies that the Hopf fibre cannot be described by a single global section, but this by itself does not establish the existence of a regular EW monopole. For an isolated magnetic monopole the corresponding asymptotic patching must also admit an extension through the full EW gauge bundle. As we show in section~\ref{sec:mono}, this additional requirement is precisely where the obstruction arises in the EWT, and it is this which excludes regular finite energy, isolated monopoles carrying non-zero asymptotic EM magnetic charge in the SM.

There is a topological current associated with the fibre curvature
\be 
K^{\mu}=-{1\over 32\pi^2}\varepsilon^{\mu\nu\rho\sigma}a_\nu f_{\rho\sigma}\,,
\ee
which obeys 
\be
\partial_\mu K^\mu=-{1\over 32\pi^2}\varepsilon^{\mu\nu\rho\sigma}\left[f_{\rho\sigma}\partial_\mu a_\nu+a_\nu\partial_\mu f_{\rho\sigma}\right]=-{1\over 32\pi^2}{\tilde f}_{\mu\nu}f^{\mu\nu}\,,
\ee
where ${\tilde f}^{\mu\nu}=\half\varepsilon^{\mu\nu\rho\sigma}f_{\rho\sigma}$ and we have used $\varepsilon^{\mu\nu\rho\sigma}\partial_\nu f_{\rho\sigma}=0$. This  leads to the Hopf invariant~\cite{Hopf1931,Whitehead1947} 
\be 
Q_H=-{1\over 32\pi^2}\int d^3{\bf x}\,\varepsilon^{ijk}a_if_{jk}\,.
\label{hopfinv}
\ee
This is a linking number similar to magnetic helicity - and also the Abelian Chern-Simons number. These are the current and charge that are associated with the homotopy group $\pi_3(S^2)={\mathbb{Z}}$. Physically it will correspond to separated monopoles/anti-monopoles as we will discuss in section~\ref{sec:mmbar}. 

\section{Gauge kinetic terms}
\label{sec:gauge}

\subsection{Cho's restricted gauge theory}
\label{sec:cho}

It is natural to split the gauge field into a component which is in the direction of ${\bf{\hat n}}$ and one which is orthogonal to it. We will discuss the most natural way to do this in the next section which is similar, but not entirely equivalent to Cho's restricted gauge theory~\cite{Cho1980}. This involves writing the $SU(2)$ gauge fields as
\be{\bf W}_\mu={\widehat{\bf W}}_\mu+{\bf {\widetilde X}}_\mu\,,
\ee
where the valence field, ${\bf {\widetilde X}}_\mu$, satisfies ${\bf {\widetilde X}}_\mu\cdot{\hat{\bf n}}=0$. From this we can write the field strength tensor ${\bf W}_{\mu\nu}={\widehat{\bf W}}_{\mu\nu}+{\bf {\widetilde X}}_{\mu\nu}$ where ${\bf{\widehat W}}_{\mu\nu}=\partial_\mu{\bf{\widehat W}}_\nu-\partial_\nu{\bf{\widehat W}}_\mu-g{\widehat{\bf W}}_\mu\times{\widehat{\bf W}}_\nu$ and ${\bf {\widetilde X}}_{\mu\nu}={\hat D}_\mu{\bf {\widetilde X}}_\nu-{\hat D}_\nu{\bf {\widetilde X}}_\mu-g{\bf {\widetilde X}}_\mu\times{\bf {\widetilde X}}_\nu$ and the ``restricted'' covariant derivative ${\hat D}_\mu$ is defined by 
${\hat D}_\mu{\bf {\widetilde X}}_\nu=\partial_\mu{\bf {\widetilde X}}_\nu-g{\widehat{\bf W}}_\mu\times{\bf {\widetilde X}}_\nu$. 

The ``restricted'' gauge field ${\widehat{\bf W}}_\mu$ is defined by ${\hat D}_\mu{\hat{\bf n}}\equiv 0$, that is $\partial_\mu{\hat{\bf n}}=g{\widehat{\bf W}}_\mu\times{\hat{\bf n}}$ and hence 
\be
{\widehat{\bf W}}_\mu=w_\mu{\bf{\hat n}}+{1\over g}{\hat{\bf n}}\times\partial_\mu{\hat{\bf n}}\,,\quad {\widehat{\bf W}}_{\mu\nu}=\left(w_{\mu\nu}-{1\over g}f_{\mu\nu}\right){\hat{\bf n}}\,,
\ee
where $w_\mu={\hat{\bf n}}\cdot{\widehat{\bf W}}_\mu$,  $w_{\mu\nu}=\partial_\mu w_\nu-\partial_\nu w_\mu$ and $f_{\mu\nu}$ is the fibre curvature described in the previous section. This choice selects the part of the $SU(2)$ connection that preserves the local Higgs-defined direction ${\hat{\bf n}}$ under parallel transport. This is the same as the split used in section~\ref{sec:noberry} where we described how it is inconsistent to ignore the fibre phase in the decomposition of the scalar kinetic term. An SU(2) gauge field has 3 colour $\times$ 4 Lorentz = 12 degrees of freedom. The Cho variables are $w_\mu$ which has 4 components and ${\bf {\widetilde X}}_\mu$ has 8, so the split as described here has the same number of degrees of freedom. At this level Cho's split is legitimate. Numerous  attempts have been made to extend it beyond this point and these usually require careful treatment of constraints, redundancies and degree-of-freedom counting which can break the gauge invariance, that would have consequences for the experimental coincidence with the EWT. Notable amongst these is Cho's own attempt to rewrite the EWT in terms of the Hopf fibration~\cite{Cho:2001gc}, but without the fibre phase.

\subsection{EM, Z and W field strength tensors}

In this section we will describe how one can define the field strength tensors for the EM, Z and W gauge fields. Using a slightly different split of the gauge fields to the previous section ${\bf W}_\mu=w_\mu{\bf{\hat n}}+{\bf X}_{\mu}$, where ${\bf X}_\mu\cdot{\bf{\hat n}}=0$ and note that ${\bf X}_\mu={\widetilde{\bf X}}_\mu+{1\over g}{\hat{\bf n}}\times\partial_\mu{\bf{\hat n}}\ne{\widetilde{\bf X}}_\mu$. With these definitions (\ref{azdefn}) we choose to write the field strength tensors for the EM and Z fields to be 
\be 
A_{\mu\nu}=\sin\theta_Ww_{\mu\nu}+\cos\theta_WY_{\mu\nu}+\sin\theta_W\,{\hat{\bf n}}\cdot{\bf X}_{\mu\nu}\,,\quad Z_{\mu\nu}=\cos\theta_Ww_{\mu\nu}-\sin\theta_WY_{\mu\nu}+\cos\theta_W\,{\bf{\hat n}}\cdot{\bf X}_{\mu\nu}\,,
\ee
where ${\bf X}_{\mu\nu}=\partial_\mu{\bf X}_\nu-\partial_\nu{\bf X}_\mu-g{\bf X}_\mu\times{\bf X}_\nu$. Using these expressions, it is simple to write the gauge kinetic term part of the EW Lagrangian as 
\be
{\cal L}_{\rm gauge}=-{1\over 4}{\bf W}_{\mu\nu}\cdot{\bf W}^{\mu\nu}-{1\over 4}Y_{\mu\nu}Y^{\mu\nu}=-{1\over 4}A_{\mu\nu}A^{\mu\nu}-{1\over 4}Z_{\mu\nu}Z^{\mu\nu}-{1\over 4}{\cal P}^{ab}W_{\mu\nu}^aW^{b\mu\nu}\,,
\label{smkinlag}
\ee
where ${\cal P}^{ab}=\delta^{ab}-{\hat n}^a{\hat n}^b$ is a projection operator. Using this we can write ${\cal P}^{ab}W_{\mu\nu}^aW^{b\mu\nu}=({\cal P}^{ac}W_{\mu\nu}^a)({\cal P}^{cb}W^{b\mu\nu})$, so we see that (\ref{smkinlag}) can be written as kinetic terms for the EM, Z and W fields. 

We can express 
\be 
{\hat{\bf n}}\cdot{\bf X}_{\mu\nu}={1\over g}\left[{\bf{\hat n}}\cdot\partial_\mu{\bf{\hat n}}\times\partial_\nu{\bf{\hat n}}-{\bf{\hat n}}\cdot D_\mu{\bf{\hat n}}\times\ D_\nu{\bf{\hat n}}\right]=-{1\over g}\left[f_{\mu\nu}+{\bf{\hat n}}\cdot D_\mu{\bf{\hat n}}\times\ D_\nu{\bf{\hat n}}\right]\,.
\ee
and, if we define $K_{\mu\nu}={\hat{\bf n}}\cdot  D_\mu{\bf{\hat n}} \times D_\nu{\bf{\hat n}}$ which is gauge invariant, then this allows to deduce that~\cite{Tornkvist:1998sm} 
\begin{eqnarray}
A_{\mu\nu}&=&\partial_\mu A_\nu-\partial_\nu A_\mu-{\sin\theta_W\over g}\left[f_{\mu\nu}+K_{\mu\nu}\right]\,,\cr
Z_{\mu\nu}&=&\partial_\mu Z_\nu-\partial_\nu Z_\mu-{\cos\theta_W\over g}\left[f_{\mu\nu}+K_{\mu\nu}\right]\,,
\end{eqnarray}
both of which are also gauge invariant. When $K_{\mu\nu}=0$, the gauge invariant EM and Z field tensors are equivalent to those defined in ref.~\cite{tHooft1974} for the Georgi-Glashow model. The key thing to observe is that they include the fibre curvature, that is, there is a geometrical part to the field strength tensor. We can also write 
\be 
G_{\mu\nu}^a={\cal P}^{ab}W_{\mu\nu}^b={\hat D}_\mu{\widetilde X}^a_\nu-{\hat D}_\nu {\widetilde X}_\mu^a\,.
\ee
where ${\hat D}_\mu$ is the derivative used in the ``restricted'' gauge theory. It will be seen below that this is the field strength tensor for the W bosons.
 
\subsection{Deriving Maxwell's equations}
\label{sec:maxwell}

In this section we will demonstrate how the classical EM sector emerges from the EW along with the interaction with the W and Z bosons with the aim of isolating the possible effects of monopoles which come from the fibre curvature. We will ignore the fermion sector and hence we will be expecting to find the sourceless Maxwell equations, modulo the effects of the W and Z bosons, and the monopoles.

The equations of motion for the gauge fields, ignoring the fermionic contributions to the currents, are 
\begin{eqnarray}
\partial_\mu Y^{\mu\nu}&=&J_Y^{\nu}=\frac{1}{2}ig^{\prime}\left[\Phi^{\dag}D^{\nu}\Phi-(D^{\nu}\Phi)^{\dag}\Phi\right]\,,\cr
D_\mu W^{\mu\nu a} &=&J_W^{\nu a}=\frac{1}{2}ig\left[\Phi^{\dag}\sigma^a(D^{\nu}\Phi)-(D^{\nu}\Phi)^{\dag}\sigma^a\Phi\right]\,,
\end{eqnarray}
where $D_\lambda W^a_{\mu\nu} = \partial_\lambda W^a_{\mu\nu} - g\varepsilon^{abc}W_\lambda^b W^c_{\mu\nu}$. Using the Hopf fibration variables that we have developed in the previous sections, we find that 
\be
g^{\prime}{\hat{\bf n}}\cdot{\bf J}_W^\mu=-gJ_Y^\mu=\quarter gg^{\prime}v^2D^{\mu}\zeta\,, \quad{\bf J}_\perp^{\mu}={\cal P}{\bf J}_W^\mu=\quarter gv^2{\hat{\bf n}}\times D^\mu{\hat{\bf n}}\,,
\ee
where ${\bf J}_\perp^\mu$ is the component of the current in the direction perpendicular to ${\bf{\hat n}}$. Using these expressions we can deduce that 
\begin{eqnarray}
\partial_\mu A^{\mu\nu}&=&\sin\theta_W D_\mu{\bf{\hat n}}\cdot {\bf G}^{\mu\nu}\,,\cr
\partial_\mu Z^{\mu\nu}&=&\quarter g_Zv^2D^{\nu}\zeta+\cos\theta_WD_\mu{\bf{\hat n}}\cdot{\bf G}^{\mu\nu}\,,\cr
{\cal D}_\mu{\bf G}^{\mu\nu}&=&\quarter gv^2\,{\hat{\bf n}}\times D^\mu{\bf{\hat n}}-S^{\mu\nu}D_\mu{\bf{\hat n}}\,,
\end{eqnarray}
where ${\cal D}_\mu{\bf G}^{\mu\nu}=D_\mu{\bf G}^{\mu\nu}+(D_\mu{\hat{\bf n}}\cdot{\bf G}^{\mu\nu}){\bf{\hat n}}$ is the tangent bundle covariant derivative and $S_{\mu\nu}=\sin\theta_W\,A^{\mu\nu}+\cos\theta_W\,Z^{\mu\nu}$. If we had included the fermions, each of these equations would have included a fermionic current, which in the case of EM would include that from the electron, that is, the standard EM current - the source in Maxwell's laws.

The Bianchi identities for smooth EW fields are $D_{[\lambda}{\bf W}_{\mu\nu]}=0$ and $\partial_{[\lambda}Y_{\mu\nu]}=0$. These can be written using the Hopf variables as 
\begin{eqnarray}
\partial_{[\lambda}A_{\mu\nu]}&=&\sin\theta_WD_{[\lambda}{\bf{\hat n}}\cdot{\bf G_{\mu\nu]}}\,,\cr
\partial_{[\lambda}Z_{\mu\nu]}&=&\cos\theta_WD_{[\lambda}{\bf{\hat n}}\cdot{\bf G_{\mu\nu]}}\,,\cr
{\cal D}_{[\lambda}{\bf G}_{\mu\nu]}&=&-S_{[\mu\nu}D_{\lambda]}{\bf{\hat n}}\,.
\end{eqnarray}
Other than ignoring the fermionic currents these expressions and the equations of motion are exact. 

If we now use the Cho decomposition from section~\ref{sec:cho}, we have that ${\hat D}_\mu{\bf{\hat n}}=0$ and therefore $D_\mu{\hat{\bf n}}=-g{\widetilde{\bf X}}_\mu\times{\hat{\bf n}}$ which implies that $(D_\mu{\hat{\bf n}})^2=g^2|{\widetilde{\bf X}}_\mu|^2$ and  $K_{\mu\nu}=g^2{\hat{\bf n}}\cdot{\widetilde{\bf X}}_\mu\times{\widetilde{\bf X}}_\nu$.
In the fiducial case where ${\bf{\hat n}}={\bf{\hat z}}$ we then find that 
\be
{1\over g^2}K_{\mu\nu}={\bf{\hat n}}\cdot{\bf{\widetilde X}}_\mu\times{\bf{\widetilde X}}_\nu=W_\mu^1W_\nu^2-W_\mu^2W_\nu^1=i(W_\mu^-W_\nu^+-W_\mu^+W_\nu^-)\,,
\ee
where $W^{\pm}_\mu={1\over\sqrt{2}}(W_\mu^1\mp iW_\mu^2)$ which makes it clear that this is the W boson contribution to the field strength tensor. Using this decomposition we can write the equations governing the gauge fields as
\begin{equation}
\begin{aligned}
\partial_\mu A^{\mu\nu}&=e\,(\hat{\bf n}\times\widetilde{\bf X}_\mu)\cdot{\bf G}^{\mu\nu}\,,
&\qquad\partial_{[\lambda}A_{\mu\nu]}&=e\,(\hat{\bf n}\times\widetilde{\bf X}_{[\lambda})\cdot{\bf G}_{\mu\nu]}\,,\\\partial_\mu Z^{\mu\nu}
&={M_Z^2\over g_Z}D^\nu\zeta+g\cos\theta_W(\hat{\bf n}\times\widetilde{\bf X}_\mu)\cdot{\bf G}^{\mu\nu}\,,
&\qquad\partial_{[\lambda}Z_{\mu\nu]}&=g\cos\theta_W(\hat{\bf n}\times\widetilde{\bf X}_{[\lambda})\cdot{\bf G}_{\mu\nu]}\,,\\\hat D_\mu{\bf G}^{\mu\nu}&=
-M_W^2\widetilde{\bf X}^{\nu}
-gS^{\mu\nu}\,(\hat{\bf n}\times\widetilde{\bf X}_\mu),
&\qquad\hat D_{[\lambda}{\bf G}_{\mu\nu]}
&=-gS_{[\mu\nu}\,(\hat{\bf n}\times\widetilde{\bf X}_{\lambda]})\,,
\end{aligned}
\label{eom}
\end{equation}
where $e=g\sin\theta_W=g^{\prime}\cos\theta_W$ is the electromagnetic gauge coupling and $M_W=\half gv$, $M_Z=\half g_Zv$ are the position (and hence $v$)  dependent W and Z masses, which are the physical masses, that we will denote $m_W$ and $m_Z$, when $v=v_{\rm SM}$. The equations (\ref{eom}) provide an exact decomposition of the bosonic EW gauge equations into the massless electromagnetic direction, the massive neutral direction, and the two-dimensional charged tangent bundle perpendicular to ${\bf{\hat n}}$.

Let us introduce a local orthonormal basis $({\bf e}_1,{\bf e}_2,\hat{\bf n})$ where ${\bf e}_1\times{\bf e}_2=\hat{\bf n}$ and define ${\bf e}_\pm = {1\over\sqrt{2}} ({\bf e}_1\pm i{\bf e}_2)$ which satisfy $\hat{\bf n}\times{\bf e}_\pm = \mp i{\bf e}_\pm$. We can, therefore, write ${\widetilde{\bf X}}_\mu=W^+_\mu {\bf e}_+ + W^-_\mu {\bf e}_-$. There is a local freedom in the choice of basis in the tangent plane that is related to the unbroken $U(1)_{\rm EM}$ symmetry. Under the rotation ${\bf e}_1\rightarrow \cos\chi\,{\bf e}_1+\sin\chi\,{\bf e}_2$ and ${\bf e}_2\rightarrow-\sin\chi\,{\bf e}_1+\cos\chi\,{\bf e}_2$ and equivalently ${\bf e}_\pm\rightarrow e^{\mp i\chi}{\bf e}_\pm$ and $W^\pm_\mu\rightarrow e^{\pm i\chi}W^\pm_\mu$, that is $W^+_\mu$ and $W^-_\mu$ transform with opposite charges under rotations of the tangent plane. 

We can define the connection associated with the tangent basis by $\omega_\mu = {\bf e}_1\cdot\partial_\mu{\bf e}_2$. Under the local rotation of the tangent basis $\omega_\mu\rightarrow\omega_\mu-\partial_\mu\chi$ and hence one finds $\widehat D_\mu{\bf e}_+=
i(\omega_\mu+gw_\mu){\bf e}_+$ and $
\widehat D_\mu{\bf e}_-=-i(\omega_\mu+gw_\mu){\bf e}_-$. 
It follows that
\begin{equation}
\widehat D_\mu\widetilde{\bf X}_\nu=({\cal D}_\mu W^+_\nu){\bf e}_+
+({\cal D}_\mu W^-_\nu){\bf e}_-\,,
\end{equation}
where ${\cal D}_\mu W^\pm_\nu= \left[\partial_\mu\pm i(\omega_\mu+gw_\mu)\right]W^\pm_\nu$. The corresponding charged field strengths are $G^\pm_{\mu\nu}={\cal D}_\mu W^\pm_\nu-
{\cal D}_\nu W^\pm_\mu$.

There is a close connection between $\omega_\mu$ and the fibre connection introduced in section 3. The curvature of $\omega_\mu$ is
\begin{equation}
\omega_{\mu\nu}=\partial_\mu\omega_\nu-\partial_\nu\omega_\mu=
\hat{\bf n}\cdot(\partial_\mu\hat{\bf n}\times\partial_\nu\hat{\bf n})=-f_{\mu\nu}\,.
\end{equation}
Therefore, locally $\omega_\mu$ and $-a_\mu$ can differ only by a
gradient. The fibre curvature can be interpreted as
the curvature associated with parallel transport of vectors in the
tangent plane to the ${\mathbb C} P^1\simeq S^2$ base.

We can now rewrite the exact equations (\ref{eom}) in terms of the
charged tangent fields using
\be
{\bf G}_{\mu\nu}=G^+_{\mu\nu}{\bf e}_++G^-_{\mu\nu}{\bf e}_-\,,
\ee
where, for real $SU(2)$ gauge fields, $W^-_\mu=(W^+_\mu)^*$ and
$G^-_{\mu\nu}=(G^+_{\mu\nu})^*$. Since $\hat{\bf n}\times\widetilde{\bf X}_\mu=-iW^+_\mu{\bf e}_+ +iW^-_\mu{\bf e}_-$, we have that 
\be
(\hat{\bf n}\times\widetilde{\bf X}_\mu)\cdot{\bf G}^{\mu\nu}
=i\left(W^-_\mu G^{+\mu\nu}-W^+_\mu G^{-\mu\nu}\right)\,.
\ee
This allows equations of motion to be written as
\begin{equation}
\begin{aligned}
\partial_\mu A^{\mu\nu}&=ie\left(W^-_\mu G^{+\mu\nu}-W^+_\mu G^{-\mu\nu}\right)\,,\\
\partial_\mu Z^{\mu\nu}&={M_Z^2\over g_Z}D^\nu\zeta+ig\cos\theta_W
\left(W^-_\mu G^{+\mu\nu}-W^+_\mu G^{-\mu\nu}\right)\,,\\{\cal D}_\mu G^{\pm\mu\nu}&=-M_W^2W^{\pm\nu}\pm igS^{\mu\nu}W^\pm_\mu\,.
\end{aligned}
\label{eompm}
\end{equation}
and the Bianchi identities become
\begin{equation}
\begin{aligned}
\partial_{[\lambda}A_{\mu\nu]}&=ie\left(W^-_{[\lambda}G^+_{\mu\nu]}-W^+_{[\lambda}G^-_{\mu\nu]}
\right)\,,\\\partial_{[\lambda}Z_{\mu\nu]}&=ig\cos\theta_W\left(W^-_{[\lambda}G^+_{\mu\nu]}-W^+_{[\lambda}G^-_{\mu\nu]}\right)\,,\\{\cal D}_{[\lambda}G^\pm_{\mu\nu]}&=\pm igS_{[\mu\nu}W^\pm_{\lambda]}\,.
\end{aligned}
\label{bianchipm}
\end{equation}
The additional step relative to (\ref{eom}) has been to complexify the two-dimensional tangent space perpendicular to $\hat{\bf n}$. Thus, $W^\pm_\mu$ can be regarded as the two oppositely charged components of the tangent gauge field, with ${\cal D}_\mu$ the connection induced on this complexified tangent bundle. 

The ``Maxwell limit'' is defined by ignoring ${\cal O}([W_\mu^{\pm})^2]$ and hence we have that $\partial_\mu A^{\mu\nu}=0$, but the Bianchi identity  $\partial_{[\lambda}A_{\mu\nu]}=0$ will only apply when the fields are smooth. If we define ${\tilde A}^{\mu\nu}=\half\varepsilon^{\mu\nu\rho\sigma}A_{\rho\sigma}$ and similarly for ${\tilde f}^{\mu\nu}$, then we have that 
\be 
\partial_\mu A^{\mu\nu}=0\,,\quad\partial_\mu{\tilde A}^{\mu\nu}=J_M^\nu=-{\sin\theta_W\over g}\partial_\mu{\tilde f}^{\mu\nu}\,,
\ee
where $J_M^\mu$ is the magnetic current defined by the dual of the fibre curvature. We can define the magnetic charge as 
\be
Q_M=\int d^3{\bf x}\,J_M^0={\sin\theta_W\over 2g}\int_{S^2_\infty}dS_i\,\varepsilon^{ijk}\,{\hat{\bf n}}\cdot(\partial_j{\hat{\bf n}}\times\partial_k{\hat{\bf n}})={4\pi\sin\theta_W\over g}N_{\hat{\bf n}}={4\pi\sin\theta_W\over g}c_1\,.
\label{QM}
\ee
Therefore, in the presence of geometric monopoles, that is $N_{{\bf{\hat n}}}=c_1\ne 0$, or in the most simple case scalar field contributions such that ${\hat{\bf n}}={\hat {\bf r}}$, and in the absence of electric charge and current,  we have $\mathbf{\nabla}\cdot{\bf E}=0$, $\mathbf{\nabla}\cdot{\bf B}=\rho_M$, $\mathbf{\nabla}\times{\bf E}=-{\bf j}_{M}-{\partial_t{\bf B}}$ and $\mathbf{\nabla}\times{\bf B}={\partial_t {\bf E}}$ where $\rho_M$ and ${\bf j}_M$ are the magnetic charge density and current density, respectively. We note that $eQ_M/(2\pi)=2c_1\sin^2\theta_W$ which is not an integer unless $\sin^2\theta_W=\half$ or 1. Hence, for $c_1=1$ and the physical mixing angle, the geometrical contribution alone does not satisfy the naive Dirac condition based on electromagnetic coupling $e$. Equation (\ref{QM}), therefore, represents the geometrical contribution to the magnetic charge. The total asymptotic electromagnetic magnetic charge also depends on the remaining EW gauge fields, in particular the hypercharge contribution, as illustrated explicitly in section~\ref{sec:mono}.

\section{Perturbative aspects of the reformulation}
\label{sec:pert}

In section~\ref{sec:hopf} we reformulated the scalar kinetic term of EWT in terms of a vector field ${\bf{\hat n}}$ and a fibre phase $\zeta$. We have argued that this reformulation is exact if $\Phi\ne 0$ and hence we should expect that the perturbative aspects of the EWT can be reproduced in these variables. In this section we will show that this is indeed the case. In addition, we will discuss in section~\ref{sec:smeft} how the reformulation points toward the possible alternative ways of going beyond the EWT that are not as obvious in the standard formulation.

\subsection{Standard Higgs mechanism adapted to fibrated variables}
\label{sec:smhiggs}

Let us consider the Higgs field defined by $\Phi={1\over\sqrt{2}}v_{\rm SM}fe^{\half i\zeta}\xi$ and perturb around the solution $f=1+\delta f$, $\zeta={\bar\zeta}+\delta\zeta$ and $\xi={\bar\xi}+\delta\xi$. This leads to the expansion
\be
\Phi={1\over\sqrt{2}}v_{\rm SM}\left[{\bar\xi}+(\delta f+\half i\delta\zeta){\bar\xi}+\delta\xi\right]e^{\half i{\bar\zeta}}\,,
\ee
and $\xi^\dag\xi=1$ implies that ${\bar\xi}^{\dag}\delta\xi+{\delta\xi}^\dag{\bar\xi}=0$.  The standard way of expressing the Higgs field when investigating the spontaneous symmetry breaking is 
\be
\Phi={1\over\sqrt{2}}\begin{pmatrix}\phi_+\cr v_{\rm SM}+\rho+i\eta\end{pmatrix}\,,
\ee
where $\phi_+\in{\mathbb{C}}$ and $\rho,\eta\in{\mathbb{R}}$. From this we can read off that ${\bar\xi}=\begin{pmatrix}0 \cr 1\end{pmatrix}$, $\delta f= \rho/v_{\rm SM}$, $\delta\zeta=2\eta/v_{\rm SM}$ and $\delta\xi=\begin{pmatrix}\phi_+/v_{\rm SM}\cr 0\end{pmatrix}$.  This implies that the neutral Goldstone boson is in the fibre direction and the charged Goldstone bosons are part of the base ${\mathbb{C}}P^1/S^2$ directions. Using the definition ${\hat n}^a=-\xi^\dag\sigma^a\xi$, one can deduce that $\delta{\hat n}^a=-[\delta\xi^\dag\sigma^a{\bar\xi}+{\bar\xi}^\dag\sigma^a\delta\xi]$ which leads
\be
\delta{\hat n}^a={2\over v_{\rm SM}}\begin{pmatrix}-{\rm Re}(\phi_+)\cr {\rm Im}(\phi_+)\cr 0\end{pmatrix}\,,
\ee
and the perturbation in the fibre connection is given by $\delta a_\mu=-2i{\bar\xi}^\dag\partial_\mu\delta\xi=0$.

In the unitary gauge we have that $\eta=0$ and $\phi_+=0$ so that $\delta\zeta=0$ and $\delta\xi=0$ and one just gets the standard result, up to quadratic perturbations, that 
\be
|D_\mu\Phi|^2-V(\Phi)={1\over 2}(\partial_\mu\rho)^2-\lambda v_{\rm SM}^2\rho^2+{1\over 8}g^2v_{\rm SM}^2(\varepsilon^{abc}W_\mu^b{\bar n}^c)^2+{1\over 8}g_Z^2v_{\rm SM}^2Z_\mu Z^\mu\,,
\ee
from which we can read off the masses $m_h=\sqrt{2\lambda}v_{\rm SM}$, $m_W=\half gv_{\rm SM}$ and $m_Z=\half g_Zv_{\rm SM}$. This yields the standard results for the mass spectrum of the EWT.

From a geometrical point of view, we have seen that the Hopf decomposition separates the three broken directions into two horizontal directions on the base ${\mathbb C}P^1$, which are ``eaten'' by $W^\pm_\mu$, and the vertical fibre direction is eaten by $Z_\mu$. The unbroken electromagnetic direction is the unique gauge direction that  leaves both the base point and fibre coordinate invariant. This connects back to section~\ref{sec:noberry}.

\subsection{Scalar perturbation theory in Hopf variables}

In the previous section we have demonstrated that the standard approach to the Higgs mechanism can be expressed in the fibrated variables as we would have expected since the theory is an exact reformulation. We now explore perturbations around $v=v_{\rm SM}$, ${\hat {\bf n}}={\hat{\bf z}}$ and $\zeta=0$, ignoring the standard approach in terms of $\rho$, $\eta$ and $\phi_+$. Of course, it is possible to convert back to the standard variables using the relations we have derived.

Let us make the perturbative ansatz $v=v_{\rm SM}+h$,
\be 
{\hat{\bf n}}=\begin{pmatrix}\pi_1\cr\pi_2\cr\sqrt{1-\pi^2_1-\pi_2^2}\end{pmatrix}=\begin{pmatrix}\pi_1\cr\pi_2\cr 1-\half\pi^2\end{pmatrix}+{\cal O}(\pi^4)\,,
\ee
where $\pi^2=\pi_1^2+\pi_2^2$ and we treat $\zeta$ at first order in perturbation theory. Now define $G_i=\half v_{\rm SM}\pi_i$ and $G^\pm=\textstyle{1\over\sqrt{2}}(G_1\mp iG_2)$  which allows us to write 
\begin{eqnarray}
{v^2\over 8}|D_\mu{\hat {\bf n}}|^2&=&{1\over 2}\left(1+{h\over v_{\rm SM}}\right)^2\left\{ 
(\partial_\mu G_1-m_WW_\mu^2)^2+(\partial_\mu G_2+m_WW_\mu^1)^2+{\cal O}(3)
\right\}\cr&=&\left(1+{h\over v_{\rm SM}}\right)^2\left\{ 
(\partial_\mu G^+-im_WW_\mu^+)(\partial_\mu G^-+im_WW^{\mu-})+{\cal O}(3)
\right\}\,.\end{eqnarray}
In addition, we define ${\widetilde G}^0$ such that  $\partial_\mu {\widetilde G}^0=\half v_{\rm SM}(\partial_\mu\zeta+a_\mu)$ allowing us to write 
\begin{eqnarray}
{v^2\over 8}|D_\mu\zeta|^2={1\over 2}\left(1+{h\over v_{\rm SM}}\right)^2\left(\partial_\mu {\widetilde G}^0-m_ZZ_\mu\right)^2\,.\end{eqnarray}
This implies that in general ${\tilde G}^0$ is given by 
\be
{\widetilde G}^0=\half v_{\rm SM}\left(\zeta+\int_\Gamma a_\mu dx^\mu\right)\,,
\ee
which involves an integral around the  path $\Gamma$ and is not a single valued scalar field. Note that the non-zero curvature of the fibre connection prevents the 1-form $\partial_\mu\zeta+a_\mu$ from being exact which is connected to the Aharonov-Bohm phase. Alternatively, we can define $G_0=\half v_{\rm SM}\zeta$ and write 
\begin{eqnarray}
{v^2\over 8}|D_\mu\zeta|^2={1\over 2}\left(1+{h\over v_{\rm SM}}\right)^2\left[\left(\partial_\mu G^0-m_ZZ_\mu\right)^2+v_{\rm SM}a_\mu\left(\partial^\mu G^0-m_ZZ^\mu)+{\cal O}(4\right)
\right]\,.\end{eqnarray}

If we choose a fixed, but general, Hopf section defined by 
\be
\xi=\begin{pmatrix}-e^{-i\alpha}\sin\half\beta\cr\cos\half\beta\end{pmatrix}\,,
\ee
with $\alpha=\alpha(\theta,\phi)$ and $\beta=\beta(\theta,\phi)$ then ${\hat{\bf n}}=(\sin\beta\cos\alpha,\sin\beta\sin\alpha,\cos\beta)$, $a_\mu=-(1-\cos\beta)\partial_\mu\alpha$ and $f_{\mu\nu}=-\sin\beta(\partial_\mu\beta\partial_\nu\alpha-\partial_\nu\beta\partial_\mu\alpha)$. Therefore, we can identify $\pi_1=\sin\beta\cos\alpha$ and $\pi_2=\sin\beta\sin\alpha$ from which we can deduce that 
\begin{eqnarray}
a_\mu&=&-{1\over 2}\left(\pi_1\partial_\mu\pi_2-\pi_2\partial_\mu\pi_1\right)\,,\cr
f_{\mu\nu}&=&-\left(\partial_\mu\pi_1\partial_\nu\pi_2-\partial_\nu\pi_1\partial_\mu\pi_2\right)\,,\cr
b_i&=&{\varepsilon_{ijk}}\partial_j\pi_1\partial_k\pi_2\,,
\end{eqnarray}
where $b_i=-\half\varepsilon_{ijk}f_{jk}$ and we have ignored ${\cal O}(\pi^4)$. We see that the geometrical aspects of the formulation appear at ${\cal O}(\pi^2)$ in perturbation theory.

It is important to emphasize what this perturbative expansion tells us about the Hopf geometry. At linear order the fibre connection vanishes, and the three angular perturbations of the Higgs field are precisely the usual Goldstone modes: the two tangent directions of the base $CP^1 \simeq S^2$ correspond to $G^\pm$, while the fibre direction corresponds to $G^0$. Consequently, the quadratic theory is identical to the usual perturbative EWT, with these fields providing the longitudinal degrees of freedom of the $W^\pm_\mu$ and $Z_\mu$. The non-trivial geometry first becomes visible at second order through the fibre connection and curvature $
a_\mu$ and $f_{\mu\nu}\sim {\cal O}(\pi^2)$
and therefore it enters the Lagrangian through nonlinear interactions rather than modifying the free-particle spectrum.

There is, however, an important distinction between this local geometrical curvature and the global topology discussed in section~3. Perturbation theory about $\hat{\bf n}=\hat{\bf z}$ takes place within a single contractible Hopf coordinate patch and hence remains in the sector $c_1=0$. A configuration with $c_1\ne 0$ cannot be obtained by any small perturbation of the EW vacuum. Thus, the Hopf geometry is already encoded in ordinary perturbation theory through nonlinear interactions, whereas its non-trivial topology is intrinsically non-perturbative.
In the next section we examine how this geometrical structure appears explicitly in the gauge interactions.

\subsection{Gauge interactions}

In section~\ref{sec:maxwell} we derived the exact equations of motion for the gauge fields in the absence of fermions (\ref{eom}) before taking the ``Maxwell limit''. Ultimately, we were able to show that when $c_1\ne 0$ there will be magnetic charge and currents in the EM theory. In this section we will discuss how the gauge interactions of the EWT are encoded in these equations of motion by expanding around a perturbative vacuum where $v=v_{\rm SM}$, ${\hat{\bf n}}={\hat{\bf z}}$, $\zeta=0$ and $a_\mu=0$ as in section~\ref{sec:smhiggs}.

From the point of view of establishing the gauge interactions it is best to work at the level of the Lagrangian that gives rise to (\ref{eom}) which is given by  
\be
{\cal L}_{\rm gauge}=-{1\over4}A_{\mu\nu}A^{\mu\nu}
-{1\over4}Z_{\mu\nu}Z^{\mu\nu}+{1\over2}m_Z^2 Z_\mu Z^\mu
-{1\over2}G^+_{\mu\nu}G^{-\mu\nu}+m_W^2 W^+_\mu W^{-\mu}
\,.
\ee
We can now make contact with the usual perturbative gauge interactions.
In the perturbative vacuum ${\hat{\bf n}}={\hat{\bf z}}$ and 
we can choose
${\bf e}_1={\hat{\bf x}}$ and ${\bf e}_2={\hat{\bf y}}$, so that
$\omega_\mu=0$, and we have that ${\cal D}_\mu W^\pm_\nu=
\left[\partial_\mu\pm iV_\mu\right]W^\pm_\nu$ where $V_\mu=eA_\mu+g\cos\theta_W Z_\mu$. It is useful to define the ordinary charged field strengths
\be
F_{\mu\nu}=\partial_\mu A_\nu-\partial_\nu A_\mu\,,\quad z_{\mu\nu}=\partial_\mu Z_\nu-\partial_\nu Z_\mu\,,\quad w^\pm_{\mu\nu}=\partial_\mu W^\pm_\nu-\partial_\nu W^\pm_\mu\,,
\ee
and $Q_{\mu\nu}=W^-_\mu W^+_\nu-W^+_\mu W^-_\nu$. This allows us to deduce that
\be
A_{\mu\nu}=F_{\mu\nu}-ieQ_{\mu\nu}\,,\quad
Z_{\mu\nu}=z_{\mu\nu}-ig\cos\theta_W Q_{\mu\nu}\,,\quad
G^\pm_{\mu\nu}=w^\pm_{\mu\nu}\pm i\left(V_\mu W^\pm_\nu
-V_\nu W^\pm_\mu\right)\,.
\label{gpmpert}
\ee

We can now separate the Lagrangian, ${\cal L}_{\rm gauge}={\cal L}_2+{\cal L}_3+{\cal L}_4$, into terms that are quadratic, cubic and quartic in the gauge fields, ${\cal L}_2$, ${\cal L}_3$ and ${\cal L}_4$, respectively. 

\begin{itemize}
\item The quadratic part is
\be
{\cal L}_2
=-{1\over4}F_{\mu\nu}F^{\mu\nu}-{1\over4}z_{\mu\nu}z^{\mu\nu}+{1\over2}m_Z^2Z_\mu Z^\mu-{1\over2}w^+_{\mu\nu}w^{-\mu\nu}+m_W^2W^+_\mu W^{-\mu}\,,
\ee
which gives the standard massless photon and the massive $Z$ and $W^\pm$ fields.

\item The cubic interactions receive contributions both from the nonlinear parts of the neutral field strengths and from $V_\mu$ in the charged field strength (\ref{gpmpert}). Combining these contributions gives
\be
{\cal L}_3={\cal L}_{WWV}=i\left[V_{\mu\nu}W^{-\mu}W^{+\nu}
+V_\mu\left(w^{+\mu\nu}W^-_\nu-w^{-\mu\nu}W^+_\nu\right)\right]\,,
\ee
where $V_{\mu\nu}=\partial_\mu V_\nu-\partial_\nu V_\mu$, and substituting for $V_\mu$ gives $g_{WW\gamma}=e$ and $g_{WWZ}=g\cos\theta_W=e\cot\theta_W$ which are the standard $WW\gamma$ and $WWZ$ three-point gauge interactions.

\item The quartic interactions follow from terms quadratic in
$V_\mu$ and $Q_{\mu\nu}$ which can be written compactly as
\be
{\cal L}_4=V_\mu V_\nu W^{+\nu}W^{-\mu}
-V_\mu V^\mu W^+_\nu W^{-\nu}+{g^2\over4}Q_{\mu\nu}Q^{\mu\nu}\,,
\label{L4gauge}
\ee
where we have used $e^2+g^2\cos^2\theta_W=g^2$. Expanding $ V_\mu=eA_\mu+g\cos\theta_W Z_\mu$ gives
\begin{eqnarray}
{\cal L}_{WW\gamma\gamma}
&=&e^2\left[A_\mu A_\nu W^{+\nu}W^{-\mu}
-A_\mu A^\mu W^+_\nu W^{-\nu}\right]\,,\cr
{\cal L}_{WW\gamma Z}
&=&eg\cos\theta_W\left[(A_\mu Z_\nu+Z_\mu A_\nu)
W^{+\nu}W^{-\mu}-2A_\mu Z^\mu W^+_\nu W^{-\nu}
\right]\,,\cr {\cal L}_{WWZZ}&=&g^2\cos^2\theta_W\left[Z_\mu Z_\nu W^{+\nu}W^{-\mu}-Z_\mu Z^\mu W^+_\nu W^{-\nu}\right]\,,\cr
{\cal L}_{WWWW}&=&\half g^2\left[(W^+_\mu W^{+\mu})(W^-_\nu W^{-\nu})-(W^+_\mu W^{-\mu})^2\right]\,.
\end{eqnarray}
\end{itemize}
Therefore, we have recovered all the conventional three- and four-point gauge interactions: $WW\gamma$ and $WWZ$ at cubic order, and $WW\gamma\gamma$, $WW\gamma Z$, $WWZZ$ and $WWWW$ at quartic
order. There are no cubic or quartic interactions containing only $A_\mu$ and $Z_\mu$. In the Hopf variables this has a particularly simple interpretation: the neutral gauge fields act as connections on the charged tangent fields $W^\pm_\mu$, while the nonlinear contributions to the neutral field strengths themselves are also generated by pairs of $W^\pm_\mu$ fields.

\subsection{Hopf formulation of effective field theory}
\label{sec:smeft}

The discussion above raises the question of whether the standard choice of global Higgs field space is itself experimentally established. The reformulation developed in this paper is exactly equivalent to the EWT for $\Phi\ne0$. At fixed $v$ the four real components of the Higgs doublet conventionally define an $S^3$,
which in the variables developed here is described by the Hopf fibration $S^1\longrightarrow S^3\longrightarrow S^2$. The base is described by ${\hat{\bf n}}\in S^2$ and the fibre by $\zeta\in S^1$, with the non-trivial relation between them encoded in the fibre connection.

Perturbative measurements about the EW vacuum probe only a local neighbourhood of a particular point in this field space and, hence, the local interactions of the fields close to vacuum. They  do not by themselves determine the global topology of the vacuum manifold. In particular, one can imagine a theory whose global angular field space has a different completion, for example $S^2\times S^1$, provided that its local geometry and interactions near the EW vacuum reproduce those of the EWT to the accuracy that has been measured. Such a theory would not be globally equivalent to the EWT, even though its perturbative expansion could agree with it to any prescribed order.

This suggests a different way of constructing an effective theory. Rather than taking the complex doublet $\Phi$ as the fundamental variable and constructing all possible higher dimensional operators from it, we can take the geometrically adapted variables $v$, ${\hat{\bf n}}$ and $\zeta$ as the starting point. Gauge
invariant operators can then be constructed from
\be
v,\quad \partial_\mu v,\quad
D_\mu{\hat{\bf n}},\quad D_\mu\zeta,\quad
A_{\mu\nu},\quad Z_{\mu\nu},\quad
G^\pm_{\mu\nu}\,,
\ee
together with their covariant derivatives.  We can also expect modifications to ${\cal L}_{\rm gauge}$, for example, the hypercharge term would allow a free function 
\be
{\cal L}_{\rm hyper}=-{1\over 4}Z_Y(v)Y_{\mu\nu}Y^{\mu\nu}\,,
\ee
with $Z_Y(v_{\rm SM})=1$~\cite{KimmYoonCho2015,EllisMavromatosYou2016}.

At the zero and two derivative level a general scalar kinetic Lagrangian consistent with the symmetries can, for example, be written in the form
\be
{\cal L}_{\rm scalar}={1\over2}F_v(v)(\partial_\mu v)^2
+{1\over8}F_n(v)(D_\mu{\hat{\bf n}})^2+{1\over8}F_\zeta(v)(D_\mu\zeta)^2-{\hat V}(v)\,,
\label{hopfeft2}
\ee
where ${\hat V}(v_{\rm SM}+\delta v)=V(v_{\rm SM}+\delta v)+{\cal O}[(\delta v)^5]$ is a potential that agrees with the EWT for perturbations below order four. The EWT corresponds to the particular choice $F_v(v)=1$ and $F_n(v)=F_\zeta(v)=v^2$. An effective theory need not retain these functional forms. For example, one could have
\be
F_n(v)=v^2\left(1+c_n{v^2\over\Lambda^2}+\cdots\right),
\qquad
F_\zeta(v)=v^2\left(1+c_\zeta{v^2\over\Lambda^2}+\cdots\right),
\ee
so that physics above the scale $\Lambda$ can modify the geometry of the base and fibre independently.

An example of this idea has recently been proposed whereby the target space of the Higgs field is ${\mathbb C}P^2$ rather than ${\mathbb C}^2$~\cite{Manton2025}. The EW $U(2)$ is identified with the isotropy group of a point in ${\mathbb C}P^2$, and locally in the neighbourhood of this point the scalar field has the form of the usual complex Higgs doublet. The constant radius orbits of $U(2)$ are ``squashed'' $S^3$ manifolds, which can themselves be regarded as Hopf circle bundles over $S^2$. Close to this point in ${\mathbb C}P^2$ the EWT is recovered, while the curvature of the full $\mathbb{C}P^2$ target produces higher dimensional corrections. This provides a concrete example in which a theory with a different global scalar geometry reproduces the local EWT and is naturally interpreted as an effective extension of it. This model provides a specific choice of the functions in (\ref{hopfeft2}). Writing the inhomogeneous ${\mathbb C}P^2$ coordinate in our variables, the Fubini--Study kinetic
term gives
\be
F_v(v)={1\over(1+v^2/(2M^2))^2},\qquad
F_n(v)={v^2\over1+v^2/(2M^2)},\qquad
F_\zeta(v)={v^2\over(1+v^2/(2M^2))^2}.
\ee
The ratio $F_\zeta/F_n=(1+v^2/(2M^2))^{-1}$ is the square of the ratio of the Hopf fibre radius to the radius of the $S^2$ base, and is, therefore, precisely the squashing of the constant-$v$ $S^3$s in the ${\mathbb C}P^2$ geometry. In the limit
$M\rightarrow\infty$ one recovers
$F_v=1$ and $F_n=F_\zeta=v^2$, as required for the EWT. At higher order in derivatives there are further geometrically natural invariants, for example
\be
[(D_\mu{\hat{\bf n}})^2]^2,\quad
(D_\mu{\hat{\bf n}}\cdot D_\nu{\hat{\bf n}})
(D^\mu{\hat{\bf n}}\cdot D^\nu{\hat{\bf n}})\,,\quad (D_\mu\zeta)^4\,,\quad
(D_\mu\zeta)^2(D_\nu{\hat{\bf n}})^2\,,\quad\dots
\ee
together with operators involving the EM, $Z$ and charged tangent
field strengths.

An important difference from the conventional expansion is that ${\hat{\bf n}}$ and $\zeta$ are dimensionless fields. Consequently canonical mass dimension is not necessarily the most natural ordering principle in this approach. A derivative expansion, similar
to that used in non-linear sigma models, is more natural. If the Hopf variables are merely regarded as a change of variables in the EWT, all powers of $v$ required to reproduce the conventional canonical dimensions must of course be retained. If instead they
are regarded as fundamental fields of an effective theory, simple operators in the Hopf variables can correspond to non-polynomial functions of the original Higgs doublet and hence to an infinite series of conventional higher dimensional interactions. This distinction is important. Requiring the theory to extend analytically through $v=0$ and to possess the global topology of the original Higgs doublet strongly restricts the allowed operators. Relaxing this requirement gives a larger class of effective theories which have the same local gauge symmetry and can have the same perturbative EW limit, but different global geometry and potentially different non-perturbative sectors.

Just as an illustration, we can calculate some of the SMEFT basis in terms of the Hopf variables. The basic identities required for the conversion are
\be
\Phi^\dagger\Phi={1\over2}v^2,\,\quad
\Phi^\dagger\sigma^a\Phi=-{1\over2}v^2{\hat n}^a\,,\quad\Phi^\dagger D_\mu \Phi
={1\over2}v\partial_\mu v +{i\over4}v^2D_\mu\zeta\,.
\ee
There are 3 operators containing only scalar fields which are given by~\cite{GrzadkowskiEtAl2010}
\begin{eqnarray}
Q_{\Phi}&=&(\Phi^\dagger\Phi)^3={1\over8}v^6\,,\cr
Q_{\Phi\Box}&=&(\Phi^\dagger\Phi)\Box(\Phi^\dagger\Phi)={1\over4}v^2\Box v^2\simeq-v^2(\partial_\mu v)^2\,,\cr
Q_{\Phi D}&=&(\Phi^\dagger D_\mu\Phi)^*(\Phi^\dagger D^\mu\Phi)
={1\over4}v^2(\partial_\mu v)^2+{1\over16}v^4(D_\mu\zeta)^2\,,
\end{eqnarray}
where $\simeq$ denotes equality up to a total derivative. Already we see that all of this, and more, are included in (\ref{hopfeft2}).

\section{Hopf fibration and fermions}
\label{sec:fermion}

In the previous sections we have shown that the bosonic Lagrangian of the EWT can be written in terms of the fibrated variables $\zeta$ and ${\hat{\bf n}}$. We will now show that the same
geometrical variables also provide a natural description of the fermionic sector. We will concentrate on a single family of up and
down quarks. The extension to leptons and additional generations is straightforward and will be discussed below.

The relevant fields are the left-handed doublet
\be
q_L=\left(\begin{array}{c}u_L\\d_L\end{array}\right)\,,
\ee
and the right-handed singlets $u_R$ and $d_R$. Their weak hypercharges are $y_q=\textstyle{1\over3}$, $y_u=\textstyle{4\over3}$, $y_d=-\textstyle{2\over3}$
which satisfy $y_u=y_q+1$, $y_d=y_q-1$.
The fermion kinetic terms are
\be
{\cal L}_{\rm kin,F}=\bar q_L i\gamma^\mu D_\mu q_L+\bar u_Ri\gamma^\mu D_\mu u_R+\bar d_Ri\gamma^\mu D_\mu d_R\,,
\ee
where
\be
D_\mu q_L=\left(\partial_\mu+{1\over2}ig' y_qY_\mu+{1\over2}ig{\bf W}_\mu\cdot{\boldsymbol\sigma}\right)q_L\,,
\ee
and
\be
D_\mu u_R=\left(\partial_\mu+{1\over2}ig'y_uY_\mu\right)u_R\,,
\quad D_\mu d_R=\left(\partial_\mu+{1\over2}ig'y_dY_\mu\right)d_R\,.
\ee
The Yukawa interactions are
\be
{\cal L}_Y=-f_u\bar q_L\widetilde\Phi u_R-f_d\bar q_L\Phi d_R+{\rm h.c.}\,,
\ee
where $\widetilde\Phi=i\sigma_2\Phi^*$.

Using the Hopf decomposition $\Phi=\textstyle{1\over\sqrt{2}}ve^{\half i\zeta}\xi$, we define the spinor orthogonal to $\xi$ by $\xi_\perp=i\sigma_2\xi^*$~\footnote{Note that this is distinct from the $\xi_\perp$ defined in section~\ref{sec:phase}.} with $\xi^\dag\xi_\perp=0$ so that $\widetilde\Phi ={1\over\sqrt{2}}ve^{-\half i\zeta}\xi_\perp$. The two spinors $(\xi_\perp,\xi)$ therefore define a local orthonormal frame in the weak doublet space. It is also  useful to define $U_\xi=(\,\xi_\perp\ \ \xi\,)\in SU(2)$. We can now expand the left-handed fermion doublet in this moving frame,
\be
q_L=u_L\xi_\perp+d_L\xi=U_\xi\begin{pmatrix}u_L\cr d_L\end{pmatrix}\,.
\label{ql}
\ee
Here, $u_L$ and $d_L$ denote the components with respect to the
Higgs-defined frame. They coincide with the usual components in
unitary gauge.

Using (\ref{ql}) the Yukawa Lagrangian becomes
\be
{\cal L}_Y=-{v\over\sqrt{2}}\left[f_u e^{-\half i\zeta}\bar u_Lu_R
+f_d e^{\half i\zeta}\bar d_Ld_R\right]+{\rm h.c.}\,,
\ee
and the remaining factors of $\zeta$ can themselves be absorbed into
the fermion fields. We now define
\be
{\cal Q}_L =e^{-\half iy_q\zeta}\begin{pmatrix}u_L\cr d_L\end{pmatrix},
\qquad{\cal Q}_R=\begin{pmatrix}e^{-\half iy_u\zeta}u_R\\
e^{-\half iy_d\zeta}d_R\end{pmatrix}\,.
\ee
The two entries of ${\cal Q}_R$ are, of course, still independent
weak singlets; writing them as a column is merely convenient
notation. One can use the fact that $y_u-y_q=-(y_d-y_q)$ to write ${\cal L}_Y$ in a particularly simple form
\be
{\cal L}_Y=-\left[\bar{\cal Q}_L M_Q(v){\cal Q}_R+\bar{\cal Q}_R M_Q^\dag(v){\cal Q}_L\right]=-\begin{pmatrix}\bar{\cal Q}_L &\bar{\cal Q}_R\end{pmatrix}{\cal M}_Q\begin{pmatrix}{\cal Q}_L\cr{\cal Q}_R\end{pmatrix}\,,
\label{yukawaH}
\ee
where
\be
M_Q={v\over\sqrt{2}}\begin{pmatrix}f_u&0\cr0&f_d\end{pmatrix}\,,\quad{\cal M}_Q=\begin{pmatrix}0&M_Q\cr M_Q^\dag&0\end{pmatrix}\,.
\ee
Thus, in the fibrated variables the Yukawa interaction is simply
a radial-field-dependent mass matrix; all dependence on the Hopf
fibre coordinate has been transferred into the fermion kinetic
terms. This redefinition also makes the connection with electric charge particularly transparent. If we define
$Q_u=\textstyle{1\over2}(y_q+1)=\textstyle{1\over2}y_u=\textstyle{2\over3}$,
$Q_d=\textstyle{1\over2}(y_q-1)=\textstyle{1\over2}y_d=-\textstyle{1\over3}$. We will see that these are precisely the charges with which the fermions transform under changes of the local Hopf frame.

For the right-handed fields one immediately obtains
\be
{\cal D}^{R}_\mu{\cal Q}_R=\begin{pmatrix}
\partial_\mu+iQ_uc_\mu&0\cr0&\partial_\mu+iQ_dc_\mu\end{pmatrix}{\cal Q}_R\,,
\ee
where $c_\mu$ is defined in section~\ref{sec:hopf}. For the left-handed fields there is an additional contribution because the Higgs-defined frame varies in space. Using the decomposition of the gauge fields developed in section~\ref{sec:gauge}, one finds
\be
U_\xi^\dag\left(\partial_\mu+{1\over2}ig{\bf W}_\mu\cdot
{\boldsymbol\sigma}\right)U_\xi=-{i\over2}(a_\mu-gw_\mu)\sigma_3
+{ig\over\sqrt{2}}\begin{pmatrix}0&W^+_\mu\cr W^-_\mu&0\end{pmatrix}\,,
\label{DR}
\ee
where $w_\mu$ is defined in section~\ref{sec:cho} and $W^\pm_\mu$ are the same charged tangent fields introduced in section~\ref{sec:maxwell}. The derivatives of the left-handed dressed fields can be written entirely in terms of the fibrated variables as
\be
{\cal D}^{L}_\mu{\cal Q}_L=
\begin{pmatrix}\partial_\mu+iQ_uc_\mu-{i\over2}D_\mu\zeta&
{ig\over\sqrt{2}}W^+_\mu\cr{ig\over\sqrt{2}}W^-_\mu&
\partial_\mu+iQ_dc_\mu+{i\over2}D_\mu\zeta
\end{pmatrix}{\cal Q}_L\,.
\label{DL}
\ee
Equations (\ref{DR}) and (\ref{DL}) are the fermionic analogue of the decomposition of the bosonic gauge sector given in section~\ref{sec:gauge}. They separate the neutral fibre connection from the charged tangent directions.

As a check, consider the perturbative vacuum in unitary gauge where
$\zeta=0$, $a_\mu=0$ and ${\hat{\bf n}}={\hat{\bf z}}$. In this
case $c_\mu=g'Y_\mu=eA_\mu-g_Z\sin^2\theta_W Z_\mu$ and $D_\mu\zeta=-g_ZZ_\mu$. Therefore, the left-handed derivative becomes
\be
{\cal D}^{L}_\mu=\begin{pmatrix}\partial_\mu+ieQ_uA_\mu
+ig_Z\left({1\over2}-Q_u\sin^2\theta_W\right)Z_\mu
&{ig\over\sqrt{2}}W^+_\mu\cr {ig\over\sqrt{2}}W^-_\mu&
\partial_\mu+ieQ_dA_\mu-ig_Z\left({1\over2}+Q_d\sin^2\theta_W\right)Z_\mu\end{pmatrix}\,,
\ee
while for the right-handed fields
\be
{\cal D}^{R}_\mu=\begin{pmatrix}\partial_\mu+ieQ_uA_\mu
-ig_ZQ_u\sin^2\theta_W Z_\mu&0\cr 0&
\partial_\mu+ieQ_dA_\mu-ig_ZQ_d\sin^2\theta_W Z_\mu\end{pmatrix}\,.
\ee
These are precisely the standard photon, $Z$ and $W^\pm$
couplings to the quarks.

There is a geometrical interpretation of the electric
charges. The local redundancy $\xi\rightarrow e^{-i\lambda/2}\xi$ and $\zeta\rightarrow\zeta+\lambda$ implies
$\xi_\perp\rightarrow e^{i\lambda/2}\xi_\perp$, $U_\xi\rightarrow U_\xi e^{i\lambda\sigma_3/2}$ and $c_\mu\rightarrow c_\mu+\partial_\mu\lambda$.
This means that the dressed fermion fields transform according to
\be
{\cal Q}_u\rightarrow e^{-iQ_u\lambda}{\cal Q}_u\,,
\qquad
{\cal Q}_d\rightarrow e^{-iQ_d\lambda}{\cal Q}_d\,.
\ee
Thus, the local $U(1)$ freedom associated with the choice of Hopf
frame acts on the dressed fermions with precisely their electric
charges. In this sense the conventional electromagnetic charge
assignment is encoded geometrically in the way that the
fermions are attached to the base and fibre of the Hopf
decomposition.

The construction for leptons is identical. Writing
\be
\ell_L=\nu_L\xi_\perp+e_L\xi,
\ee
with $y_\ell=-1$ and $y_e=-2$, $Q_\nu=\textstyle{1\over2}(y_\ell+1)=0$ and $Q_e=\textstyle{1\over2}(y_\ell-1)=-1$.
The electron Yukawa coupling  takes the same radial form
as (\ref{yukawaH}), while the neutrino remains massless in the
minimal EWT. Additional generations can be included by promoting the
Yukawa couplings to matrices in flavour space. The usual flavour
mixing then appears in the charged $W^\pm$ interactions after the
mass matrices have been diagonalized.

Finally, the fermionic construction has a possible global
significance. In a sector with $c_1\ne0$ there is no single
globally defined choice of $\xi$. If two Hopf patches are related
on their overlap by $\xi_i=e^{-i\lambda_{ij}/2}\xi_j$ and $\zeta_i=\zeta_j+\lambda_{ij}$, 
then the corresponding fermions are patched according to ${\cal Q}_{u,i}=e^{-iQ_u\lambda_{ij}}{\cal Q}_{u,j}$ and ${\cal Q}_{d,i}=e^{-iQ_d\lambda_{ij}}{\cal Q}_{d,j}$.
The fermions can therefore be regarded as sections of the Hopf $U(1)$ bundle, with their electric charges
determining the corresponding representation. For quarks, a full
discussion of the global patching also involves the colour sector
and the global form of the EW gauge group, which we have not
included here.  

\section{Are there monopole solutions?}
\label{sec:mono}

One of the main motivations for writing the EWT using the Hopf fibration is the study of monopoles. As well as rewriting the fields in terms of ${\bf{\hat n}}$ and $\zeta$ we have explained that the first Chern number, which quantifies the geometry, is equivalent to the monopole number and we have also shown how monopoles can, in principle, source Maxwell's equations. In this section we will investigate whether there can be actual monopole solutions. 

\subsection{Boundary conditions on $S^2_\infty$}

An important issue in the construction of monopole solutions is the boundary conditions. We will search for solutions which are maps from $S_{\infty}^2\rightarrow {\cal M}$ where ${\cal M}$ is the vacuum manifold of the EWT. In the ${\mathbb C}^2$ formulation of the EWT we have that ${\cal M}=S^3$ and hence it is the boundary conditions on $S_\infty^2$ that we need to fix.

The key observation is that, in order to construct a finite-energy solution, we want to impose $D_i\Phi=0$ at $r=\infty$.
If we define ${\cal W}_i=L_i+\frac{1}{2}igW_i^{a}U^{\dag}_{\rm L}\sigma^aU_{\rm L}$ and the projection operator ${\cal P}^{ab}=\delta^{ab}-{\bar n}^a{\bar n}^b$, then the part of the energy density which is quadratic in the fields (that is, the mass matrix) is given by 
\begin{eqnarray}
{\cal E}_{\rm mass}=\frac{1}{8}v^2\left[g^2{\cal X}_i^a{\cal X}_i^a+(g^2+g^{\prime 2}){\cal Z}_i^2\right]
\end{eqnarray}
where 
\be
{\cal A}_i=\sin\theta_{\rm W}{\cal W}_i^a{\bar n}^a+\cos\theta_{\rm W}Y_i\,,\quad {\cal Z}_i=\cos\theta_{\rm W}{\cal W}_i^a{\bar n}^a-\sin\theta_{\rm W}Y_i\,,
\ee
are the photon and $Z$ boson fields with $\tan\theta_{\rm W}=g^{\prime}/g$ and ${\cal X}_i^a={\cal P}^{ab}{\cal W}_i^b$ with ${\bar n}^a{\cal X}_i^a=0$ 
Therefore, we impose ${\cal\chi}_i^a\equiv 0$ and ${\cal Z}_i\equiv 0$ - that is we set the energy in the directions associated with $W^\pm$ and $Z$ to be zero at $r=\infty$. In addition, it is also necessary to impose a condition in the direction associated with the photon, $A_i^\infty$. In this case we find that 
\begin{eqnarray}
Y_i&=&{\sec\theta_W\,A_i^\infty+\tan\theta_W\,{\bar n}^a{\cal R}^{ab}L_i^b\over 1+\tan^2\theta_W\,{\bar n}^a{\cal R}^{ab}{\bar n}^b}\,,\cr
W_i^a&=&{\cal R}^{ab}\left[\tan\theta_W\,Y_i{\bar n}^b-L_i^b\right]\,.
\end{eqnarray}
If we set $g^{\prime}=0$ then $\tan\theta_W=0$ and $\sec\theta_W=1$ and therefore $Y_i=A_i^{\infty}$ and $W_i^a=-{\cal R}^{ab}L_i^b$ which implies that
\be
 {\bf W}^\infty_{i}=-{1\over g}\left(\partial_i\gamma\,{\bf {\hat G}}+\sin\gamma\,\partial_i{\bf{\hat G}}+(\cos\gamma-1)\,{\bf{\hat G}}\times\partial_i{\hat{\bf G}}\right)\,.
 \ee

Let us now write $\alpha={\bar n}^a{\cal R}^{ab}{\bar n}^b$ and $\beta_i={\bar n}^a{\cal R}^{ab}L_i^b$ then $Y_i=(\sec\theta_W\,A_i^\infty+\tan\theta_W\,\beta_i)/(1+\alpha\tan^2\theta_W)$ and $W_i^a{\bar n}^a=\alpha\tan\theta_W\,Y_i-\beta_i$. Therefore, we have that 
\begin{eqnarray}
\label{zi}
Z_i&=&(\alpha-1)\sin\theta_W\,Y_i-\cos\theta_W\,\beta_i={(\alpha-1)\tan\theta_W\,A_i^\infty-\sec\theta_W\,\beta_i\over 1+\alpha\tan^2\theta_W}\,.
\end{eqnarray}

Equation \ref{zi} makes explicit that, for $g'\neq0$, the finite-energy condition $Z_i=0$ fixes the asymptotic electromagnetic gauge potential in terms of the Higgs-frame geometry
\be
A_i^\infty=-{\beta_i\over (1-\alpha)\sin\theta_W}\,.
\ee
Equivalently, $Y_i=\cos\theta_W A_i^\infty$ and
$W_i^a\bar n^a=\sin\theta_W A_i^\infty$, so that the surviving asymptotic gauge field lies entirely in the unbroken $U(1)_{\rm EM}$ direction.  Notice that $\alpha=1$ when $\hat{\bf n}=\bar{\bf n}$;
in a non-trivial magnetic sector this makes the single-patch form of the connection potentially singular unless $\beta_i$ vanishes with the appropriate rate.  This is the local manifestation of the patching
structure which becomes important for monopole configurations below.

\subsection{Nambu monopole configurations}
\label{sec:nambu}

The defining properties of the Nambu monopole~\cite{Nambu1977} are that  ${\hat{\bf n}}={\hat{\bf r}}$ and ${\cal W}_i$ is set to be zero at $r=\infty$ which is equivalent to
$D_i{\hat{\bf n}}=D_i\zeta=0$. We also assume that $Y_i=0$ and $\zeta=0$. When ${\bar{\bf n}}={\hat{\bf z}}$, these assumptions lead to $\gamma=\theta$ and ${\hat{\bf G}}=-{\hat{\boldsymbol{\phi}}}$, the gauge field is given by ${\bf W}_i=h(r){\bf W}_i^\infty$ and the Higgs field is set to $v=v_{\rm SM}f(r)$. The asymptotic gauge field is given by 
\be
{\bf W}_i^\infty=-{1\over gr}\left[{\hat\phi}_i{\hat{\boldsymbol{\theta}}}-{\hat\theta}_i{\hat{\boldsymbol{\phi}}}
+{\hat\phi}_i{\hat{\bf r}}\left({1-\cos\theta\over\sin\theta}\right)\right]={1\over g}\left[{\hat{\bf r}}\times\partial_i{\hat{\bf r}}-{\hat\phi}_i{\hat{\bf r}}\left({1-\cos\theta\over r\sin\theta}\right)\right]\,.
\label{winf}
\ee
If we define
\be
M_i=M{\hat\phi}_i={\bf W}_i^\infty\cdot{\hat{\bf z}}=\left({1-\cos\theta\over gr\sin\theta}\right){\hat\phi}_i={\tan\half\theta\over gr}{\hat\phi}_i\,,
\label{MM}
\ee
then we have that ${\bf W}_i^\infty={1\over g}{\hat{\bf n}}\times\partial_i{\hat{\bf n}}-M_i{\hat{\bf r}}$ where 
\be
\nabla\times{\bf M}={1\over r\sin\theta}{\partial\over\partial\theta}\left[M\sin\theta\right]{\hat{\bf r}}-{1\over r}{\partial\over\partial r}[rM]{\hat{\boldsymbol{\theta}}}={{\hat{\bf r}}\over gr^2}\,,
\ee
which illustrates that it is the configuration $M_i$ that leads to the monopole magnetic field. This is a configuration with $N_{\bf{\hat n}}=1$, but it is not spherically symmetric: the Z-flux is all forced along the $\theta=\pi$ direction which is the well-documented Z-string.

One can construct higher charge configurations by constructing a  unit vector from a rational function $R(z)=p(z)/q(z)$ where $p$ and $q$ are coprime polynomials of the complex coordinate $z=e^{i\phi}\tan\half\theta$ for the unit $S^2$ (see for example~\cite{Donaldson:1984ugy,Houghton:1997kg}). The standard form is 
\be
{\hat{\bf n}}={1\over 1+|R|^2}\left(R+R^*,-i(R-R^*),1-|R|^2\right)\,.
\ee
Using a procedure similar to that above, one can construct a field with $\tan\half\gamma=|R|$ and 
\be
{\hat{\bf G}}=-{1\over 2|R|}\left(i(R-R^*),R+R^*,0\right)\,,\qquad
U_L={1\over \sqrt{1+|R|^2}}\begin{pmatrix}1 & -R^*\cr R & 1\end{pmatrix}\,,
\ee
and a field configuration 
\be
\Phi={v_{\rm SM}\over\sqrt{2}}f(r){1\over\sqrt{1+|R|^2}}\begin{pmatrix}-R^*\cr 1\end{pmatrix}\,.
\label{ratfield}
\ee

One can calculate the asymptotic gauge field directly for a
general rational map. If we define $R=|R|e^{i\alpha}$ then $
{\hat{\bf n}}=(\sin\gamma\cos\alpha,\sin\gamma\sin\alpha,\cos\gamma)$.
For the field configuration  (\ref{ratfield}), the fibre connection is
\be
a_i=i{R^*\partial_iR-R\partial_iR^*\over1+|R|^2}
=-(1-\cos\gamma)\partial_i\alpha\,.
\label{ai}
\ee
The conditions $D_i{\hat{\bf n}}=0$ and $D_i\zeta=0$ imply that 
the components of the $SU(2)$ gauge field parallel to and perpendicular to ${\hat{\bf n}}$ are given by $g{\bf W}^{\infty}_i\cdot{\hat{\bf n}}=a_i$ and $g{\bf W}^{\infty}_{i\perp} ={\hat{\bf n}}\times\partial_i{\hat{\bf n}}$.
Hence, for a general rational map,
\be
{\bf W}^{\infty}_i={1\over g}{\hat{\bf n}}\times\partial_i{\hat{\bf n}}+{a_i\over g}{\hat{\bf n}}={1\over g}{\hat{\bf n}}\times\partial_i{\hat{\bf n}}-{\widetilde M}_i{\hat{\bf n}}\,,
\label{ratW}
\ee
where
\be
{\widetilde M}_i=-{a_i\over g}=-{i\over g}{R^*\partial_iR-R\partial_iR^*\over1+|R|^2}={1-\cos\gamma\over g}\partial_i\alpha\,.
\label{ratM}
\ee
For the choice ${\bar{\bf n}}={\hat{\bf z}}$ used here one can also
verify that ${\widetilde M}_i={\bf W}^{\infty}_i\cdot{\hat{\bf z}}=M_i$, so that (\ref{ratM}) is the direct generalization of (\ref{MM}). Since ${\tilde M}_i=-a_i/g$, its curvature is proportional to the pullback of the area form under the rational map, and its total flux is $\Phi_M=\Phi_A/\sin\theta_W=4\pi N/g$, where $N$ is the degree of $R$.

The case of $R(z)=z$ clearly reproduces the standard case of ${\hat{\bf n}}={\hat{\bf r}}$ and $\alpha=\phi$ so that (\ref{ratW}) and (\ref{ratM}) become (\ref{winf}) and (\ref{MM}).  When $R(z)=z^2$ we have that 
\be 
\Phi={v_{\rm SM}\over 2}f(r){1\over\sqrt{1+\cos^2\theta}}\begin{pmatrix}-(1-\cos\theta)e^{-2i\phi}\cr 1+\cos\theta\end{pmatrix}\,,
\ee
which leads to 
\be 
{\hat{\bf n}}={1\over 1+\cos^2\theta}\left((1-\cos^2\theta)\cos 2\phi,(1-\cos^2\theta)\sin 2\phi, 2\cos\theta\right)\,,
\ee
and 
\be 
{\hat{\bf G}}=\begin{pmatrix}\sin 2\phi\cr -\cos 2\phi\cr 0\end{pmatrix}\,.
\ee
Moreover, we also have that $\alpha=2\phi$ and, hence, the fibre connection and the corresponding ${\tilde M}_i$ are given by
\be
a_i=-2(1-\cos\gamma)\partial_i\phi=-{2(1-\cos\theta)^2\over r\sin\theta(1+\cos^2\theta)}\,{\hat\phi}_i\,,
\ee
and
\be
{\tilde M}_i={2(1-\cos\theta)^2\over gr\sin\theta(1+\cos^2\theta)}\,{\hat\phi}_i\,.
\ee

To write the full $SU(2)$ field explicitly, define the two unit
vectors tangent to the target $S^2$,
\be
{\hat{\bf e}}_\gamma=(\cos\gamma\cos2\phi,\cos\gamma\sin2\phi,-\sin\gamma)\,,\qquad{\hat{\bf e}}_\alpha=(-\sin2\phi,\cos2\phi,0)\,,
\ee
which satisfy ${\hat{\bf n}}\times{\hat{\bf e}}_\gamma
={\hat{\bf e}}_\alpha$ and ${\hat{\bf n}}\times{\hat{\bf e}}_\alpha
=-{\hat{\bf e}}_\gamma$\,.
Since
\be
\partial_i{\hat{\bf n}}={1\over r}{d\gamma\over d\theta}
{\hat\theta}_i{\hat{\bf e}}_\gamma+{2\sin\gamma\over r\sin\theta}
{\hat\phi}_i{\hat{\bf e}}_\alpha\,,
\ee
we can write (\ref{ratW}) entirely in terms of
$\theta$ as
\be
{\bf W}^{\infty}_i={2\over gr(1+\cos^2\theta)}
\left[\sin\theta\,{\hat\theta}_i{\hat{\bf e}}_\alpha
-\sin\theta\,{\hat\phi}_i{\hat{\bf e}}_\gamma-{(1-\cos\theta)^2\over\sin\theta}{\hat\phi}_i{\hat{\bf n}}\right]\,.
\ee
Hence, the rational map determines not only the angular Higgs
configuration, but also the compensating asymptotic gauge field
required by ${\cal W}_i^\infty=0$. 

For a rational-map configuration the unit vector ${\hat{\bf n}}$ and the fibre curvature $f_{ij}$ can remain smooth, whereas the fibre connection $a_i$, and hence ${\tilde M}_i=-a_i/g$, is necessarily singular in a single coordinate patch whenever the map has non-zero degree, that is, when $c_1\ne 0$. For $R=z^N$ the singularity lies on the negative $z$-axis and behaves as
\be
{\tilde M}_i\sim {2N\over gr(\pi-\theta)}\,{\hat\phi}_i\,,
\ee
while for a general rational map the singularities occur along the directions associated with the poles of $R$, with multiplicity equal to their order. These are patch, or string, singularities rather than singularities of the curvature itself, and in the Nambu interpretation they correspond to $Z$-strings carrying the winding away from the monopole. Thus, rational maps can provide natural higher-charge or multi-string boundary data, but they do not by themselves give finite-energy isolated monopole solutions.

\subsection{Cho-Maison monopole configuration}

The Cho-Maison (CM) monopole~\cite{ChoMaison1997,KimmYoonCho2015,ZhangZouCho2020}  has  $Z_i=0$ at $r=\infty$ which requires that $Y_i=\cot\theta_W\left(W_i\cdot {\bar{\bf n}}\right)$ there. If one uses the definition of $M_i$ in (\ref{MM}) then $Y_i=\cot\theta_WM_i$ at $r=\infty$ and  one finds that $A_i=M_i/\sin\theta_W$. If one defines the flux for a field X to be 
\be 
\Phi_X=\int dS_i\epsilon_{ijk}\partial_jX_k\,,
\ee
then we deduce that $\Phi_A=4\pi/e$ where $e=g\sin\theta_W$ for the CM monopole over a spherical surface of radius $R\rightarrow\infty$ and $\Phi_Z=0$. It is a ``true'' magnetic monopole with $g_m=e\Phi_A/(2\pi)=2$. In addition to the flux due to a geometrical monopole discussed in section~\ref{sec:maxwell} there is an additional component due to the hypercharge $4\pi\cos^2\theta_W/e$ which makes the total $4\pi/e$.

The Higgs field configuration ansatz for the CM monopole is 
\be 
\Phi=\frac{v_{\rm SM}}{\sqrt{2}}f(r)\begin{pmatrix}
\cos\half\theta & -e^{-i\phi}\sin\half\theta \cr e^{i\phi}\sin\half\theta& 
\cos\half\theta\end{pmatrix}\begin{pmatrix}0\cr 1\end{pmatrix}=\frac{v_{\rm SM}}{\sqrt{2}}f(r)\begin{pmatrix}
-e^{-i\phi}\sin\half\theta\cr
\cos\half\theta\end{pmatrix}\,,
\ee
which has ${\hat{\bf n}}={\hat{\bf r}}$. The key issue is how to choose the gauge fields which need to have the property that $Y_i=\cot\theta_WM_i$ at $\infty$, so that $Z_i=0$ and $A_i^\infty=M_i/\sin\theta_W$. From the earlier discussion at $r=\infty$  we have that 
\be 
{\bf W}_i={\bf W}^{\infty}_i+\tan\theta_W\,Y_i\,{\hat{\bf n}}={1\over g}{\bf{\hat r}}\times\partial_i{\bf{\hat r}}+(\tan\theta_W\,Y_i-M_i){\hat{\bf r}}={1\over g}{\bf{\hat r}}\times\partial_i{\bf{\hat r}}\,.
\ee
Therefore, the gauge configuration for the CM is 
\be
{\bf W}_i={h(r)\over g}{\hat {\bf r}}\times\partial_i{\bf{\hat r}}=h(r){{\hat\theta}_i{\boldsymbol{\hat\phi}}-{\hat\phi}_i{\boldsymbol{\hat\theta}}\over gr}\,,\quad Y_i=\cot\theta_WM_i\,,
\ee
for which 
\begin{eqnarray}
Z_i&=&\cos\theta_W({\bf W}_i\cdot{\bar{\bf n}})-\sin\theta_WY_i=\cos\theta_W\left[
{h(r)\over g}{\bar{\bf n}}\cdot{\hat{\bf r}}\times\partial_i{\bf{\hat r}}-M_i\right]\,,\cr
A_i&=&\sin\theta_W({\bf W}_i\cdot{\bar{\bf n}})+\cos\theta_WY_i={1\over\sin\theta_W}\left[\sin^2\theta_W
{h(r)\over g}{\bar{\bf n}}\cdot{\hat{\bf r}}\times\partial_i{\bf{\hat r}}+\cos^2\theta_WM_i\right]\,.
\end{eqnarray}
Note that ${\bf{\widehat W}}_i=h(r)\left[{\bf W}_i^\infty+M_i{\hat{\bf r}}\right]$.

In order to calculate the energy functional we will use the Hopf fibrated version of the energy. We have that $a_i=-gM_i$  and ${\bf W}_i\cdot{\hat{\bf n}}=0$ so that $D_i\zeta=a_i+g^\prime Y_i=0$. We also have that $g{\bf W}_i\times{\bf {\hat n}}=h\partial_i{\bf{\hat n}}$ and therefore
\be
E_{\rm gauge}=4\pi\int_0^\infty dr\left[
{(h^\prime)^2\over g^2}
+{h^2(2-h)^2\over 2g^2r^2}\right]+E_Y\,,\quad E_Y={2\pi\over (g^{\prime})^2}\int_0^\infty{dr\over r^2}\,,
\label{enCM}
\ee
and $E_Y$ is clearly divergent at $r=0$.

The rational-map construction of the previous section can also be applied to the CM configuration. The boundary condition is obtained, as in the $N=1$ case above, by choosing $Y_i^\infty=\cot\theta_W M_i$. The additional longitudinal $SU(2)$ component then cancels the ${\widetilde M}_i{\hat{\bf n}}$ term and the asymptotic fields become
\be
\Phi={v_{\rm SM}\over\sqrt{2}}f(r)
{1\over\sqrt{1+|R|^2}}
\begin{pmatrix}-R^*\\1\end{pmatrix},
\qquad
{\bf W}_i={h(r)\over g}{\hat{\bf n}}\times\partial_i{\hat{\bf n}}\,,
\qquad
Y_i=\cot\theta_W {\widetilde M}_i\,,
\ee
where these expressions should be regarded as initial conditions for a numerical relaxation rather than an exact ansatz when the rational map is not spherically symmetric. If the rational map has degree $N$, the corresponding asymptotic
fluxes are
$\Phi_A=4\pi N/e$, $\Phi_Y=4\pi N/ g'$ and $\Phi_Z=0$.
Note that the rational map changes the angular distribution of the magnetic field, but does not remove the CM hypercharge
obstruction; in particular, the non-zero hypercharge flux persists for any $N\ne0$. We will return to this issue in section~\ref{sec:nonsph}.

\subsection{Spherically symmetric monopole solutions?}
\label{sec:proof}

The CM configuration considered in the previous section
has infinite-energy due to the $U(1)_Y$ contribution, and we have argued that this could be true in general. However, this calculation uses a restricted spherical ansatz for the gauge fields and does not by itself exclude the possibility that additional spherically symmetric degrees of freedom could regularize the solution. In this section we consider a more general spherically symmetric ansatz and show that this, in fact, is not possible. Similar conclusions were reached in ref.~\cite{GervalleVolkov2022}.

We will concentrate  on the $c_1=1$ sector, fixing ${\hat{\bf n}}={\hat{\bf r}}$, write $v=v_{\rm SM}f(r)$ and $\zeta=\zeta(r)$. We will also use the standard Hopf section (\ref{Nambu}). The most general static, purely magnetic, spherically symmetric $SU(2)_L$ gauge field can be written in the form~\cite{ForgacsManton1980}
\be
{\bf W}_i={1\over g}\left[h(r){\hat{\bf n}}\times\partial_i{\hat{\bf n}}+k(r)\partial_i{\hat{\bf n}}+{\ell(r)\over r}{\hat{\bf n}}\,{\hat r}_i\right]\,.
\label{genW}
\ee
The first term is the CM ansatz, $k(r)$ is the coefficient of the second component in the tangent plane to $S^2_{\hat n}$ and $\ell(r)$ is that for the radial component parallel to ${\hat{\bf n}}$. The corresponding spherically symmetric hypercharge field, in the northern coordinate patch, is
\be
Y_i =\cot\theta_W M_i+{y(r)\over g^{\prime}r}{\hat r}_i\,.
\label{genY}
\ee
The coefficient of the angular part is fixed by the magnetic
topological sector\footnote{This will ultimately be seen to be the obstruction to a finite-energy monopole solution.}, whereas $y(r)$ is an arbitrary radial component. Therefore, the spherically symmetric ansatz contains the five radial functions
$f,h,k,\ell,y$, together with the radial phase $\zeta$.

Using this ansatz we find that 
\begin{eqnarray}
D_i{\hat{\bf n}}&=&(1-h)\partial_i{\hat{\bf n}}+k\,{\hat{\bf n}}\times\partial_i{\hat{\bf n}}\,,\cr D_i\zeta&=&{1\over r}\left[r\zeta'+y-\ell\right]{\hat r}_i={q(r)\over r}{\hat r}_i\,.
\end{eqnarray}
The $SU(2)_L$ field strength can also be calculated directly from
(\ref{genW}). Defining
\be
{\cal A}=h'-{\ell k\over r},\quad{\cal B}=k'-{\ell(1-h)\over r},\quad{\cal C}=2h-h^2-k^2\,,
\ee
one obtains
\begin{eqnarray}
g{\bf W}_{r\theta}={\cal A}{\hat{\boldsymbol\phi}}
+{\cal B}{\hat{\boldsymbol\theta}}\,,\quad g{\bf W}_{r\phi}
=\sin\theta\left[-{\cal A}{\hat{\boldsymbol\theta}}
+{\cal B}{\hat{\boldsymbol\phi}}\right]\,,\quad
g{\bf W}_{\theta\phi}={\cal C}\sin\theta\,{\hat{\bf r}}\,,\quad
g'Y_{\theta\phi}=\sin\theta\,,\nonumber\\
\end{eqnarray}
and $Y_{r\theta}=Y_{r\phi}=0$. Notice that the radial function $y(r)$ does not appear in the hypercharge field strength and is locally a pure gauge degree of freedom. By contrast, the angular contribution in (\ref{genY}) is fixed by the magnetic sector.

Substituting these expressions into the static EWT energy gives
\be
\begin{split}
E=4\pi\int_0^\infty dr\bigg\{&{1\over2}v_{\rm SM}^2r^2(f')^2
+{1\over4}v_{\rm SM}^2f^2\left[(1-h)^2+k^2\right]+{1\over8}v_{\rm SM}^2f^2q^2+{\lambda v_{\rm SM}^4\over4}r^2(f^2-1)^2\\&+{1\over g^2}\left[\left(h'-{\ell k\over r}\right)^2+\left(k'-{\ell(1-h)\over r}\right)^2+{(2h-h^2-k^2)^2\over2r^2}\right]\bigg\}+E_Y\,.
\end{split}
\label{genE}
\ee
The important feature of this expression is that every term is positive semi-definite including $E_Y$ which is completely independent of all the radial profile functions and divergent.

Before drawing this conclusion it is useful to show explicitly how the additional functions in (\ref{genE}) are related to gauge degrees of freedom and constraints. 
First, consider an $SU(2)_L$ transformation
\be 
U=\exp\!\left[\half i\omega\hat{\bf r}\cdot\boldsymbol{\sigma}\right],
\ee
where $\omega\equiv\omega(r)$, leaves ${\hat{\bf n}}={\hat{\bf r}}$ unchanged, but rotates the two tangent components according to $(1-h)+ik\rightarrow e^{-i\omega}[(1-h)+ik]$, while $\ell\to\ell-r\omega^\prime$ and $\zeta\to\zeta-\omega$. This freedom can be used to choose $k=0$. 

With this choice, the dependence of the energy on $y,\ell$ and
$\zeta$ is
\be
E\supset 4\pi\int_0^\infty dr\left[{1\over g^2}{\ell^2(1-h)^2\over r^2}+{1\over8}v_{\rm SM}^2f^2q^2\right]\,.
\ee
Since $y$ has no derivative term, variation with respect to $y$
gives the constraint $f^2q=0$. Away from a possible zero of the Higgs field this implies $q=0$, or equivalently $D_r\zeta=0$.
The equation obtained by varying $\ell$ then reduces to
$\ell(1-h)^2=0$. For a non-trivial monopole configuration $h$ cannot be identically equal to one; in particular regularity of the non-Abelian fields at the origin requires the monopole profile to leave the vacuum value. It follows that $\ell=0$ and $y=-r\zeta'$. A radial $U(1)_Y$ transformation can finally be used to choose $\zeta=0$ and $y=0$. We have, therefore, shown that, modulo gauge transformations and the radial constraint equations, the general spherically symmetric magnetic ansatz reduces to the CM case.

We have shown rigorously that the singularity of the CM monopole is not an artifact of having omitted additional spherically symmetric degrees of freedom. Within the static $c_1=1$ sector of the unmodified EWT, assuming spherical symmetry and the usual EW vacuum boundary conditions, the most general magnetic ansatz is gauge equivalent, after imposing its radial constraints, to the CM form and necessarily has infinite energy since it includes $E_Y$. 

So far we have restricted the discussion to purely magnetic solutions. Allowing non-zero temporal components produces the dyonic extension of the general spherical ansatz. Their electric field strengths and scalar couplings make additional positive contributions to the static energy and, therefore, cannot cancel the hypercharge term. Hence, our conclusion concerning the absence of finite-energy spherical solutions is unchanged.

\subsection{Beyond spherical symmetry}
\label{sec:nonsph}

We have shown in the previous section that the divergence of CM monopole due to $E_Y$ cannot be avoided in the case of spherical symmetry. In this section we will show that the same obstruction applies to any regular isolated configuration carrying non-zero asymptotic electromagnetic charge, independent of any symmetry assumption. At the end of the section we will discuss the various caveats, which of course must include the Nambu monopole.

Let us first consider the hypercharge field for a static spherically symmetric configuration whose magnetic field must have the form ${\bf B}_Y=B_Y(r){\hat{\bf r}}$. $Y_{\mu\nu}$  satisfies the Bianchi identity $\partial_{[\lambda}Y_{\mu\nu]}=0$ at every point where the fields are regular. For the magnetic field
this implies that 
\be
{\bf \nabla}\cdot{\bf B}_Y=
{1\over r^2}{d\over dr}\left(r^2B_Y\right)=0\,,
\ee
and hence $B_Y(r)=q_Y/r^2$, allowing us to calculate 
\be
\Phi_Y=\int_{S^2_\infty}d{\bf S}\cdot{\bf B}_Y=4\pi q_Y\,,
\ee
which is independent of $r$. The contribution to the energy from the hypercharge field is
\be
E_Y={1\over2}\int d^3x\,{\bf B}_Y^2
={\Phi_Y^2\over 8\pi}\int_0^\infty {dr\over r^2}\,,
\ee
which diverges at the origin if $q_Y\ne 0$, that is, $\Phi_Y\ne 0$. Conversely, regularity at $r=0$ requires $q_Y=0$ and hence a regular spherically symmetric configuration cannot carry hypercharge magnetic flux. 

We can make a similar argument for non-spherical hypercharge distributions. Let $\Sigma_1$ and $\Sigma_2$ be two closed surfaces which are homologous in a region where the EWT fields are regular, by which we mean that together they form the boundary of a three-dimensional volume ${\cal V}$. Since the hypercharge field strength satisfies $\nabla\cdot{\bf B}_Y=0$ in ${\cal V}$ the flux through the two surfaces must be equal
\be 
\int_{\Sigma_2}{\bf B}_Y\cdot d{\bf S}=\int_{\Sigma_1}{\bf B}_Y\cdot d{\bf S}\,,
\label{fluxsurf}
\ee
using the divergence theorem.

For a completely regular configuration in ${\mathbb R}^3$ an
enclosing surface can be continuously contracted to an arbitrarily small surface. In this limit the flux of a regular field tends to zero and then (\ref{fluxsurf}) implies $\Phi_Y(S^2_\infty)=0$ and therefore a globally regular $U(1)_Y$ gauge field cannot support a non-zero magnetic flux at infinity. This is independent of the shape of the configuration.

There is also a useful energetic version of the same argument. Suppose that a non-zero flux $\Phi_Y$ persists through a family of spheres of radius $r$, without assuming that the field itself is spherically symmetric. The Cauchy--Schwarz inequality gives
\be
\int_{S_r}dS\,|{\bf B}_Y|^2\geq {1\over4\pi r^2} \left(\int_{S_r}d{\bf S}\cdot{\bf B}_Y\right)^2={\Phi_Y^2\over4\pi r^2}\,,
\ee
from which it follows that
\be
E_Y(r<\epsilon)\geq{\Phi_Y^2\over8\pi}\int_0^\epsilon {dr\over r^2}\,,
\ee
which is divergent for any $\Phi_Y\ne0$. Thus deforming the configuration away from spherical symmetry, or concentrating the flux in some angular directions rather than others, cannot remove the ultraviolet divergence associated with a fixed hypercharge magnetic flux. 

For an isolated EW monopole, the asymptotic massive fields must vanish, and hence we have that $\Phi_Y=\cos\theta_W\Phi_A$. Consequently, any regular finite-energy object carrying a genuine long range electromagnetic magnetic charge encounters the same obstruction as the spherically symmetric CM configuration. Similar numerical conclusions were found in ref.~\cite{GervalleVolkov2023}.

There are several ways in which the assumptions of this argument can fail:
\begin{itemize}
\item A hypercharge magnetic source or a singular core could violate $\nabla\cdot{\bf B}_Y=0$, but this would require physics beyond the EWT. This could use ideas along the line of those discussed in section~\ref{sec:smeft}. In particular, one could allow a function of $v$ in front of the hypercharge kinetic as proposed in refs.~\cite{KimmYoonCho2015,EllisMavromatosYou2016}. However, we note that this does not remove the bundle obstruction itself (see section~\ref{sec:obs}).

\item The configuration need not describe an isolated monopole. For example, the magnetic flux associated with a Nambu monopole~\cite{Nambu1977} can be connected to a string, in which case an enclosing surface cannot everywhere be taken to lie in the vacuum in the same way as for an isolated monopole.

\item A monopole-antimonopole configuration~\cite{Taubes:1982ie,Taubes:1982if,Kleihaus:1999sx,Kleihaus:2000hx,Saurabh:2019rrp,Patel:2023sfm} could have $\Phi_Y(S_\infty^2)=0$, but it is likely that such a configuration would be unstable. We will discuss this type of configuration in detail in section~\ref{sec:mmbar}.

\end{itemize}
These possibilities are not counterexamples to the argument above because the relevant surfaces and boundary conditions are excluded by our assumptions, but they are illustrations of how one might want to try to construct topological configurations.

\subsection{Bundle obstruction to SM monopoles}
\label{sec:obs}
The relation to the Hopf geometry is also worth emphasizing. A configuration with $c_1\ne0$ requires non-trivial patching of the Hopf fibre over the $S^2$ base. As discussed in section~\ref{sec:geom}, this Hopf $U(1)$ bundle should be distinguished from the physical EWT gauge bundle. The projected field $\hat{\bf n}$ can, therefore, have a perfectly regular description in individual coordinate patches without this being sufficient to define a regular isolated EW monopole.

To see the additional requirement, consider a large sphere $S_R^2$ surrounding the monopole, chosen sufficiently far from the core that the fields lie in the EWT vacuum. Its interior is the 3-ball, $B_R^3=\{{\bf x}\in{\mathbb R}^3:|{\bf x}|\leq R\}$ with $\partial B_R^3=S_R^2$, and in the limit $R\rightarrow\infty$ this gives the sphere $S^2_\infty$ used above. Now cover $S_R^2$ with northern and southern patches, and  their overlap  is topologically an equatorial circle $S^1$. A monopole of integer winding number $m\in\pi_1(U(1)_{\rm EM})=\mathbb Z$ can be described by an EM transition function given by 
\be 
h_m(\phi)=\begin{pmatrix} e^{im\phi}&0\cr 0&1\end{pmatrix}\in U(1)_{\rm EM}\subset U(2)_{\rm EW}={SU(2)_L\times U(1)_Y\over\mathbb{Z}_2}\,.
\ee

Let $i:U(1)_{\rm EM}\rightarrow U(2)_{\rm EW}$ denote the inclusion of the EM subgroup into the full EW gauge group. Here, ``inclusion'' simply means that the same EM transformation is now regarded as an element of $U(2)_{\rm EW}$. This induces a map on homotopy classes of loops,
\be
i_*:\pi_1(U(1)_{\rm EM})\rightarrow\pi_1(U(2)_{\rm EW})\,,
\ee
where $i_*(m)$ is the winding of the same loop when it is allowed to explore the full EW group. The kernel of this map,
\be
\ker i_*=\{m\in\pi_1(U(1)_{\rm EM}):i_*(m)=0\}\,,
\ee
consists precisely of those EM windings which become contractible in the full EW gauge group. Since $\det h_m(\phi)=e^{im\phi}$ and $\pi_1(U(2))=\mathbb Z$, this gives $i_*(m)=m$ and $\ker i_*=0$. Hence, there are no non-zero electromagnetic windings that become contractible in the full EW group. This is a manifestation of arguments presented in a different way and not specifically in the context of the EWT in ref.~\cite{Preskill1984}.

This is directly related to regularity of the monopole core: $B_R^3$ is contractible and hence every smooth $U(2)_{\rm EW}$ gauge bundle over it is topologically trivial. A transition function on its boundary can, therefore, arise from a smooth EWT configuration throughout the interior only if it is contractible when regarded as a loop in $U(2)_{\rm EW}$. $\ker i_*=0$ implies  that this is impossible for $m\ne0$. Equivalently,
\be
\pi_2\left(U(2)_{\rm EW}/U(1)_{\rm EM}\right)
\cong \ker i_*=0\,,
\ee
which is the bundle form of the familiar result that the SM vacuum manifold has $\pi_2(S^3)=0$.

This argument is the geometrical equivalent of the same obstruction as the energetic arguments presented above. For an isolated monopole the finite-energy condition requires $Z_i\rightarrow0$ at large distance, and hence $\Phi_Y=\cos\theta_W\Phi_A$. A non-zero electromagnetic magnetic charge, therefore, requires non-zero hypercharge flux. However, a smooth $U(1)_Y$ curvature obeys the Abelian Bianchi identity and cannot carry such a flux through the contractible interior. This means that if the required singular hypercharge core is allowed, then the ordinary SM hypercharge kinetic term gives the UV divergence derived above.

 The geometrical condition $c_1\ne0$ by itself is somewhat weaker: in principle the $W^\pm$, $Z$ and fibre fields can redistribute the geometrical flux, and configurations such as the Nambu monopole evade the assumptions of an isolated monopole by remaining attached to a $Z$-string. We, therefore, distinguish the two statements: $\Phi_A(S^2_\infty)\ne0$ implies no regular isolated monopole in the unmodified SM, whereas $c_1\ne0$ implies non-trivial Hopf geometry, but not by itself the existence of an isolated monopole solution. In this sense the Hopf geometry identifies the possible monopole sectors, while the topology of the full EW gauge bundle determines whether they can be realized as regular isolated EW configurations.

\subsection{Monopole/Anti-monopole configurations}
\label{sec:mmbar}

Using the Nambu solution one can construct an ansatz for the scalar field of a monopole/anti-monopole (${M{\bar M}}$) pair including a relative rotation by $\gamma$ about an axis joining the pair. For a well separated pair one can use a product ansatz
\begin{equation}
\Phi={v_{\rm SM}\over\sqrt{2}}f(r_{\bar M},r_{M}) U_{\bar M}(\theta_{\bar M},\phi_{\bar M})U_\gamma U_M(\theta_{M},\phi_M){\bf e}\,,  
\end{equation}
where  the spherical polar coordinates $(r_M,\theta_M,\phi_M)$ are measured relative to the position of the monopole and similarly for $(r_{\bar M},\theta_{\bar M},\phi_{\bar M})$ for the anti-monopole. We choose ${\bar{\bf n}}={\hat{\bf z}}$ to be the direction joining the pair, the relative rotation matrix is given by 
\begin{equation}
U_\gamma=\cos\half\gamma\,I_2-i\sin\half\gamma\,{\hat{\bf z}}\cdot{\boldsymbol{\sigma}}=\begin{pmatrix}
    e^{-\half i\gamma}& 0\cr 
    0 & e^{\half i\gamma}
\end{pmatrix}\,.
\end{equation}
For this ansatz, the bilinear unit vector can be calculated using 
\be 
{\hat n}^a={\cal R}^{ab}(\theta,-{{\hat{\boldsymbol{\phi}}}}){\cal R}^{bc}(-\gamma,{\hat{\bf z}}){\cal R}^{cd}(\theta,-{\hat {\boldsymbol{\phi}}}){\hat z}^d
\ee
and one finds that 
\begin{eqnarray}
{\hat{\bf n}}&=&\half(1-\cos\gamma){\hat{\bf z}}+\half(1+\cos\gamma){\bf{\hat n}}_{\rm sph}+\sin\gamma\sin\theta{\boldsymbol{\hat\phi}}\cr
&=&\cos\theta\,{\hat{\bf r}}+\sin\theta\cos\gamma\,{\hat{\boldsymbol{\theta}}}+\sin\theta\sin\gamma\,{\hat{\boldsymbol{\phi}}}\,.
\label{mmbar}
\end{eqnarray}
There are two interesting limits of this expression: $\gamma=0$ where ${\hat{\bf n}}={\hat{\bf n}}_{\rm sph}$, so that the projected ${\bf{\hat n}}$-field coincides with that of a sphaleron configuration; and $\gamma=\pi$ where ${\hat{\bf n}}={\bf{\hat z}}$, corresponding to the vacuum. Note that because it is a $M{\bar M}$, it has $N_{\bf{\hat n}}=0$.

This configuration can be thought of as suspension\footnote{For a pointed topological space, the reduced suspension is the geometrical construction $\Sigma X=X\times[0,1]/((X\times\{0\})\cup(X\times\{1\})\cup(\{x_0\}\times[0,1]))$. The key result in the context of this paper is that $\Sigma S^n\cong S^{n+1}$.} of $S^2$ using the coordinate $\gamma$ which is denoted $\Sigma S^2 \simeq S^3$ providing the non-contractible loop which underlies the EW sphaleron~\cite{Manton1983} and  the Taubes $M{\bar M}$ construction~\cite{Taubes:1982ie}. Note that using the notation of \cite{Manton1983} $\mu=(\pi-\gamma)/2$.  The Hopf projection maps the non-trivial element of $\pi_3(S^3)$ carried by the Higgs field onto the corresponding non-trivial element of $\pi_3(S^2)$ carried by $\hat{\mathbf n}$ representing the generator of $\pi_1\!\left({\rm Maps}(S^2\rightarrow S^3)\right)\cong\pi_3(S^3)={\mathbb Z}$.

If we define
\be
z_1=-\cos\half\gamma\sin\theta\,e^{-i\phi}\,,\quad z_2=\cos\half\gamma\cos\theta+i\sin\half\gamma\,,
\end{equation}
then $|z_1|^2+|z_2|^2=1$
and hence $(z_1,z_2)$ parameterise an $S^3$
\begin{equation}
-(n_1+i n_2)=2z_1^*z_2\,,\quad  -n_3=|z_1|^2-|z_2|^2,
\end{equation}
which is the Hopf map - note that we have used the convention $n^a=-\xi^\dag\sigma^a\xi$ throughout this paper.

The topology can be visualised in terms of preimages.  The preimage of the north pole, $\hat{\mathbf n}=+\hat{\mathbf z}$, is $z_1=0$ whereas the preimage of the south pole $\hat{\mathbf n}=-\hat{\mathbf z}$, is $z_2=0$, which are linked circles in $S^3$, known as the Hopf link. Equivalently, if the suspension coordinate is represented by a monotonic radial function $\gamma=\gamma(r)$ with $\gamma(0)=-\pi$, $\gamma(\infty)=+\pi$ then the preimage of $+\hat{\mathbf z}$ is the compactified $z$ axis, while the preimage of $-\hat{\mathbf z}$ is the equatorial circle at the radius
$r_0$ satisfying $\gamma(r_0)=0$.  The two curves link once.

The same result can be expressed locally using the fibre connection, $a_i=-2i\,z^\dag\partial_i z$ where $z=\begin{pmatrix}z_1\cr z_2\end{pmatrix}$ from which we can deduce 
\be
a_i=\cos\theta\,\partial_i\gamma+\sin\theta\sin\gamma\,\partial_i\theta-(1+\cos\gamma)\sin^2\theta\,\partial_i\phi\,,
\ee
and $f_{ij}$, and hence substitute into (\ref{hopfinv}) to obtain $Q_H=1$.

It is important that $Q_H$ is an invariant of the \emph{whole} one-parameter family.  For a fixed value of $\gamma$, $\hat{\mathbf n}$ is only a map $S^2\rightarrow S^2$, and there is no Hopf invariant; it  appears only after the loop parameter, $\gamma$, is included so that the domain becomes $S^3$.

Using arguments similar to those used in the previous sections, one obtains
\be
{\mathbf W}_i(\gamma)={h(r)\over g}\left[(1+\cos\gamma)\hat{\mathbf r}\times\partial_i\hat{\mathbf r}-\sin\gamma\,\partial_i\hat{\mathbf r}\right]\,.
\ee
At the midpoint $\gamma=0$,
\begin{equation}
\mathbf W_i={2h(r)\over g}\hat{\mathbf r}\times\partial_i\hat{\mathbf r}\,,
\end{equation}
which is the Klinkhamer-Manton sphaleron gauge field for $g^\prime=0$.  At the two ends, $\gamma=\pm\pi$ and so ${\mathbf W}_i=0$, so the loop closes on vacuum configurations.

Taubes' original monopole/anti-monopole loop was formulated in a theory with an adjoint Higgs field whose vacuum manifold is $S^2$~\cite{Taubes:1982ie,Taubes:1982if} and this is classified by
\be
\pi_1\!\left({\rm Maps}(S^2\rightarrow S^2)\right)\cong
\pi_3(S^2)=\mathbb Z\,.
\ee
The non-contractible loop we have constructed here instead starts from $\pi_3(S^3)$.  The Hopf fibration $ S^1\rightarrow S^3\rightarrow S^2$ provides the bridge between the two constructions. Its long exact homotopy sequence contains
\be
\dots\rightarrow \pi_3(S^1)\rightarrow\pi_3(S^3)
\stackrel{p_*}{\rightarrow}\pi_3(S^2)\rightarrow\pi_2(S^1)\rightarrow\dots\,.
\ee
Since $\pi_3(S^1)=\pi_2(S^1)=0$ and the induced map is an isomorphism $p_*:\pi_3(S^3)\overset{\simeq}{\rightarrow}\pi_3(S^2)$,  (\ref{mmbar}) is an explicit representative of this isomorphism.

The finite-separation ${M{\bar M}}$ product ansatz is not pointwise identical to the spherically symmetric EW sphaleron. Rather, it and the Klinkhamer-Manton family are different representatives of the same non-contractible class, and their asymptotic Higgs fields have the same Hopf projection.  The min-max representative with one unstable mode is the sphaleron.

There is also an important distinction between the Hopf invariant associated with the full loop and the Chern-Simons number of its midpoint.  For a loop joining neighbouring EW vacua we have that $\Delta N_{\rm CS}=1$, whereas the sphaleron at the midpoint has $N_{\rm CS}=\half$. The appropriate topological correspondence is that ${\rm deg}(U)=Q_H[\hat{\mathbf n}]=\Delta N_{\rm CS}=1$, rather than an identification of the Hopf invariant with the instantaneous value $N_{\rm CS}=\half$ at the sphaleron. In this sense the Hopf projection gives a geometrical representation of the same topology which is usually described in terms of Higgs winding and the non-Abelian Chern-Simons number. The unit winding of the Higgs field is converted by the Hopf projection into a unit linking number of the preimages of $\hat{\mathbf n}$.

\section{Summary and conclusions}

In this paper we have shown how the major components of the EWT can be written in terms of the Higgs field $v$, a vector field ${\bf{\hat n}}\in S^2$ and $\zeta\in S^1$ illustrating the Hopf fibration. Here is a summary of the main results:
\begin{itemize}
\item We reformulated the kinetic term of the EWT in the fibrated variables and demonstrated that the phase $\zeta$ is essential in maintaining gauge invariance. This is an exact local reformulation, with the same number of degrees of freedom as the conventional description. An important point is that the fibre phase $\zeta$ and the associated fibre connection cannot in general be discarded: they are required both to preserve the number of degrees of freedom and to maintain gauge invariance. 

\item The geometrical structure becomes particularly important globally. The fibre curvature is $f_{\mu\nu}=-{\hat{\bf n}}\cdot
(\partial_\mu{\hat{\bf n}}\times\partial_\nu{\hat{\bf n}})$
and its first Chern number is equal to the winding number of
${\hat{\bf n}}$, that is, $c_1=N_{\bf{\hat n}}$. Thus, although the Hopf variables can be introduced locally, for any non-zero Higgs field, non-trivial patching of the fibre becomes unavoidable when $c_1\ne0$. Such sectors cannot be reached by perturbing within a single coordinate patch around the EW vacuum and are intrinsically non-perturbative.

\item We have also shown that this geometrical formulation gives a natural decomposition of the EW gauge sector. The photon, $Z$ boson and charged $W^\pm$ fields can be separated in a gauge invariant way, and the fibre curvature appears explicitly in the corresponding field strength tensors. In the limit in which the massive fields vanish, the resulting equations have the form of Maxwell's equations with a magnetic current determined by the non-trivial geometry of ${\hat{\bf n}}$.

\item The standard perturbative properties of the EWT could be derived, but also the reformulation of the SM points to an alternative way of defining an EFT for going beyond the EWT. We illustrated this in comparison with the usual SMEFT. Our argument was that the standard formulation of the EWT, and indeed ${\cal M}=S^3$, are specific to the choice of geometry. Other geometric formulations can lead to radically different global properties. This concept is something that has been pointed out by a number of other authors, for example ~\cite{AlonsoJenkinsManohar2016b,GripaiosRandalWilliams2018,FalkowskiRattazzi2019,CohenCraigLuSutherland2021,Manton2025,Criado2026}.  

\item The same construction extends to the fermionic sector, where the local freedom in the choice of Hopf frame acts on the dressed fermion fields with their usual electric charges. The fermionic construction, therefore, gives more than a change of variables. The Higgs field defines a local moving frame in weak-isospin space, and with respect to this frame the Yukawa interactions become purely radial and the geometrical information is transferred to the fermion kinetic terms. The residual local $U(1)$ freedom in the choice of Hopf frame acts on the dressed fermions with precisely their electromagnetic charges. Thus, the same geometry which separates the photon, $Z$ and $W^\pm$ directions in the bosonic sector also provides a geometrical interpretation of the electromagnetic representations of the fermions.

\item The principal application considered here has been the possible existence of monopoles in the EWT. The Nambu configuration realizes a non-trivial ${\hat{\bf n}}$ winding, but the magnetic flux is attached to a $Z$-string and the object is not an isolated monopole. Rational maps provide a systematic generalization to higher winding number and could provide useful boundary data for non-spherical configurations.

\item The CM construction~\cite{ChoMaison1997} instead removes the asymptotic $Z$ flux and produces a genuine long-range electromagnetic magnetic field, but at the price of a singular hypercharge field. We have shown explicitly that this divergence is not an artifact of the usual restricted CM ansatz (see also refs.~\cite{GervalleVolkov2022}).  The most general static, spherically symmetric magnetic configuration in the $c_1=1$ sector is, after imposing the constraint equations and fixing the remaining gauge freedom, gauge equivalent to the CM form and its hypercharge energy necessarily diverges at the origin. Hence, there is no regular, finite-energy, spherically symmetric, isolated monopole in the unmodified EWT.

\item More importantly, the obstruction does not depend on spherical symmetry. For a regular fundamental hypercharge field $\nabla\cdot{\bf B}_Y=0$ implying $\Phi_Y(S^2_\infty)=0$ unless there is additional UV physics or there are additional sources of Z-flux. Since finite-energy for an isolated EW monopole requires the massive $Z$ field to vanish asymptotically, we have that $\Phi_Y=\cos\theta_W\,\Phi_A$. Consequently, a non-zero long-range electromagnetic magnetic charge requires a non-zero hypercharge magnetic flux. These two requirements are incompatible with each other for a regular configuration. Conceptually, the gauge-bundle obstruction does not depend on spherical symmetry.

\item Using a product ansatz we constructed the ${\bf{\hat n}}$ and gauge fields for a monopole/anti-monopole pair. This configuration has  $N_{\bf{\hat n}}=0$ and  Hopf invariant $Q_H=1$ for the one-parameter family of loops parameterized by $\gamma$. In addition, we have explained the connection between this and the sphaleron. 
\end{itemize}

The Hopf formulation of the EWT makes these conclusions particularly transparent. Non-trivial geometry of the Higgs field is perfectly possible within the EWT, but a non-zero geometrical monopole number does not by itself imply the existence of a regular isolated monopole. The first Chern number $c_1=N_{\hat {\bf n}}$ characterizes the non-trivial patching required to lift $\hat{\bf n}$ back to the Higgs variables. The existence of a smooth physical monopole additionally requires the asymptotic $U(1)_{\rm EM}$ gauge bundle to extend through the full EW gauge theory. In the EWT it does not: a non-zero electromagnetic transition function remains non-contractible in $U(2)_{\rm EW}$.

Earlier gauge-invariant and moving-frame reformulations of the electroweak theory contain many of the local
ingredients of the construction used here.~\cite{Chernodub:2008bd,Faddeev:2008} The additional step taken in this work is to identify these variables explicitly
with the Hopf fibration and to retain the fibre connection as a geometrical object, thereby exposing the associated
Chern and Hopf invariants and allowing the global gauge-bundle obstruction to isolated Standard Model monopoles to
be formulated directly.

The Hopf fibration provides both a natural language for identifying monopole-like sectors of the SM and a simple explanation of why a finite-energy isolated magnetic monopole is so difficult to realize without new physics. There are two issues: $c_1=N_{\hat {\bf n}}$ identifies the monopole geometry, while $\ker\left[
\pi_1(U(1)_{\rm EM})\rightarrow\pi_1(U(2)_{\rm EW})\right]$ determines whether it admits a smooth isolated core. For the  SM the first quantity can be non-zero whereas the second contains only the zero-winding element. The CM hypercharge singularity and its infinite energy are, therefore, not artifacts of the particular spherical ansatz, but consequences of this more fundamental gauge-bundle obstruction. The Hopf geometry identifies the possible monopole sectors, while the topology of the full EW gauge bundle determines whether those sectors can be realized as regular isolated SM configurations. Finally, we comment that many of these issues are resolved in the Two Higgs Doublet Model (2HDM) where Hopf fibration is ``spontaneous'' in the monopole sector and all the divergences are resolved~\cite{Battye:2023cbr,Battye:2024iec,Battye:2026a,Battye:2026b}. 

\acknowledgments

We would like to thank Apostolos Pilaftsis, Dominic Viatic, Adam Thomasson, Steven Cotterill and Jeff Forshaw for collaboration on related work on the 2HDM which led us to this project. We have benefited from some interesting interactions with Nick Manton and Tanmay Vachaspati and from the use of ChatGPT (GPT-5.6 Sol).

\bibliography{hopf.bib}
\bibliographystyle{apsrev4-1}

\def\jnl#1#2#3#4#5#6{\hang{#1, {\it #4\/} {\bf #5}, #6 (#2).}}
\def\jnltwo#1#2#3#4#5#6#7#8{\hang{#1, {\it #4\/} {\bf #5}, #6; {\it
ibid} {\bf #7} #8 (#2).}} 
\def\prep#1#2#3#4{\hang{#1, #4.}} 
\def\proc#1#2#3#4#5#6{{#1, in {\it #3 (#4)\/}, edited by #5,\ (#6).}}
\def\book#1#2#3#4{\hang{#1, {\it #3\/} (#4, #2).}}
\def\jnlerr#1#2#3#4#5#6#7#8{\hang{#1 [#2], {\it #4\/} {\bf #5}, #6. {Erratum:} {\it #4\/} {\bf #7}, #8.}}
\def\prl{Phys.\ Rev.\ Lett.}
\def\pr{Phys.\ Rev.}
\def\pl{Phys.\ Lett.}
\def\np{Nucl.\ Phys.}
\def\prp{Phys.\ Rep.}
\def\rmp{Rev.\ Mod.\ Phys.}
\def\cmp{Comm.\ Math.\ Phys.}
\def\mpl{Mod.\ Phys.\ Lett.}
\def\apj{Ap.\ J.}
\def\apjl{Ap.\ J.\ Lett.}
\def\aap{Astron.\ Ap.}
\def\cqg{Class.\ Quant.\ Grav.} 
\def\grg{Gen.\ Rel.\ Grav.}
\def\mn{MNRAS}
\def\ptp{Prog.\ Theor.\ Phys.}
\def\jetp{Sov.\ Phys.\ JETP}
\def\jetpl{JETP Lett.}
\def\jmp{J.\ Math.\ Phys.}
\def\zpc{Z.\ Phys.\ C}
\def\cupress{Cambridge University Press}
\def\pup{Princeton University Press}
\def\wss{World Scientific, Singapore}
\def\oup{Oxford University Press}
\def\asj{Astron.~J}
\def\imp{Int.\ J.\ Mod.\ Phys.}

\end{document}